\documentclass[prb,showpacs,amsfonts,amssymb,amsmath,floats,twocolumn,aps]{revtex4}

\usepackage[pdftex]{graphicx}
\usepackage{dcolumn}
\usepackage{bm}
\usepackage{color}

\newcommand{\ff}[1]{{\boldsymbol #1}}

\newcommand{\tr}{\mbox{tr} \,}

\begin{document} 

\title{
Spin dynamics and relaxation in the classical-spin Kondo-impurity model beyond the Landau-Lifschitz-Gilbert equation
}

\author{Mohammad Sayad and Michael Potthoff}
\affiliation{I. Institut f\"ur Theoretische Physik, Universit\"at Hamburg, Jungiusstra\ss{}e 9, 20355 Hamburg, Germany}

\begin{abstract}
The real-time dynamics of a classical spin in an external magnetic field and locally exchange coupled to an extended one-dimensional system of non-interacting conduction electrons is studied numerically. 
Retardation effects in the coupled electron-spin dynamics are shown to be the source for the relaxation of the spin in the magnetic field. 
Total energy and spin is conserved in the non-adiabatic process. 
Approaching the new local ground state is therefore accompanied by the emission of dispersive wave packets of excitations carrying energy and spin and propagating through the lattice with Fermi velocity.
While the spin dynamics in the regime of strong exchange coupling $J$ is rather complex and  governed by an emergent new time scale, the motion of the spin for weak $J$ is regular and qualitatively well described by the Landau-Lifschitz-Gilbert (LLG) equation. 
Quantitatively, however, the full quantum-classical hybrid dynamics differs from the LLG approach. 
This is understood as a breakdown of weak-coupling perturbation theory in $J$ in the course of time. 
Furthermore, it is shown that the concept of the Gilbert damping parameter is ill-defined for the case of a one-dimensional system.
\end{abstract} 
 

%75.78.-n  Magnetization dynamics
%75.78.Jp Ultrafast magnetization dynamics and switching
%75.60.Jk Magnetization reversal mechanisms
%75.10.Hk Classical spin models
%75.10.Lp Band and itinerant models

\maketitle 

%--------------------------------------------------------------------------------------------------------------
\section{Introduction}
\label{sec:intro}
%--------------------------------------------------------------------------------------------------------------

The Landau-Lifshitz-Gilbert (LLG) equation \cite{LL,LLG,1353448} has originally been considered to describe the dynamics of the magnetization of a macroscopic sample.
Nowadays it is frequently used to simulate the dynamics of many magnetic units coupled by exchange or magnetostatic interactions, i.e., in numerical micromagnetics. \cite{Aharoni}
The same LLG equation can be used on an atomistic level as well. \cite{TKS08,SHNE08,BMS09,FI11,EFC+14}
For a suitable choice of units and for several spins $\ff S_{m}(t)$ at lattice sites $m$, it has the following structure:
\begin{eqnarray}
  \frac {d \ff S_{m}(t)} {dt}
  &=&
  \ff S_{m}(t) \times \ff B
  +
  \sum_{n} J_{mn}
  \ff S_{m}(t)  \times \ff S_{n}(t)
  \nonumber \\
  &+&
  \sum_{n}
  \alpha_{mn} \ff S_{m}(t) \times 
  \frac {d \ff S_{n}(t)} {dt} \: .
\label{eq:llg}
\end{eqnarray}
It consists of precession terms coupling the spin at site $m$ to an external magnetic field $\ff B$ and, via exchange couplings $J_{mn}$, to the spins at sites $n$.
Those precession terms typically have a clear atomistic origin, such as the Ruderman-Kittel-Kasuya-Yoshida (RKKY) interaction \cite{RK54,Kas56,Yos57} which is mediated by the magnetic polarization of conduction electrons.
The non-local RKKY couplings $J_{mn} = J^{2} \chi_{mn}$ are given in terms of the elements $\chi_{mn}$ of the static conduction-electron spin susceptibility and the local exchange $J$ between the spins and the local magnetic moments of the conduction electrons.
Other possibilities comprise direct (Heisenberg) exchange interactions, intra-atomic (Hund's) couplings as well as the spin-orbit and other anisotropic interactions.
The relaxation term, on the other hand, is often assumed as local, $\alpha_{mn} = \delta_{mn} \alpha$, and represented by purely phenomenological Gilbert damping constant $\alpha$ only.
It describes the angular-momentum transfer between the spins and a usually unspecified heat bath.

On the atomistic level, the Gilbert damping must be seen as originating from microscopic couplings of the spins to the conduction-electron system (as well as to lattice degrees of freedom which, however, will not be considered here). 
There are numerous studies where the damping constant, or tensor, $\alpha$ has been computed numerically from a more fundamental model including electron degrees of freedom explicitly \cite{ON06,BNF12,UMS12} or even from first principles. \cite{AKvSH95,AKH+96,KK02,CG03,EMKK11,Sak12}
All these studies rely on two, partially related, assumptions:
(i) The spin-electron coupling $J$ is weak and can be treated perturbatively to lowest order, i.e., the Kubo formula or linear-response theory is employed. 
(ii) The classical spin dynamics is slow as compared to the electron dynamics. 
These assumptions appear as well justified but they are also {\em necessary} to achieve a simple effective spin-only theory by eliminating the fast electron degrees of freedom.

The purpose of the present paper is to explore the physics beyond the two assumptions (i) and (ii). 
Using a computationally efficient formulation in terms of the electronic one-particle reduced density matrix, we have set up a scheme by which the dynamics of classical spins coupled to a system of conduction electrons can be treated numerically exactly. 
The theory applies to arbitrary coupling strengths and does not assume a separation of electron and spin time scales.
Our approach is a quantum-classical hybrid theory \cite{Elz12} which may be characterized as Ehrenfest dynamics, similar to exact numerical treatments of the dynamics of nuclei, treated as classical objects, coupled to a quantum system of electrons 
(see, e.g., Ref.\ \onlinecite{MH00} for an overview). 
Some other instructive examples of quantum-classical hybrid dynamics have been discussed recently. \cite{Daj14,FLE14}

The obvious numerical advantage of an effective spin-only theory, as given by LLG equations of the form (\ref{eq:llg}), is that in solving the equations of motion there is only the time scale of the spins that must be taken care of.
As compared to our hybrid theory, much larger time steps and much longer propagation times can be achieved. 
Opposed to ab-initio approaches \cite{AKvSH95,AKH+96,MPC14} we therefore consider a simple one-dimensional non-interacting tight-binding model for the conduction-electron degrees of freedom, i.e., electrons are hopping between the nearest-neighboring sites of a lattice. 
Within this model approach, systems consisting of about 1000 sites can be treated easily, and we can access sufficiently long time scales to study the spin relaxation. 
An equilibrium state with a half-filled conduction band is assumed as the initial state.
The subsequent dynamics is initiated by a sudden switch of a magnetic field coupled to the classical spin. 
The present study is performed for a {\em single} spin, i.e., we consider a classical-spin Kondo-impurity model with antiferromagnetic local exchange coupling $J$, while the theory itself is general and can be applied to more than a single or even to a large number of spins as well.

As compared to the conventional (quantum-spin) Kondo model, \cite{Kon64,Hew93} the model considered here does not account for the Kondo effect and therefore applies to situations where this is absent or less important, such as for systems with large spin quantum numbers $S$, strongly anisotropic systems or, as considered here, systems in a strong magnetic field. 
To estimate the quality of the classical-spin approximation {\em a priori} is difficult. \cite{SGP12,GL14,DLZFR15}
For one-dimensional systems, however, a quantitative study is possible by comparing with full quantum calculations and will be discussed elsewhere. \cite{SRP15}

There are different questions to be addressed:
For dimensional reasons, one should expect that linear-response theory, even for weak $J$, must break down at long times. 
It will therefore be interesting to compare the exact spin dynamics with the predictions of the LLG equation for different $J$.
Furthermore, the spin dynamics in the long-time limit can be expected to be sensitively dependent on the low-energy electronic structure. 
We will show that this has important consequences for the computation of the damping constant $\alpha$ and that $\alpha$ is even ill-defined in some cases.
An advantage of a full theory of spin and electron dynamics is that a precise microscopic picture of the electron dynamics is available and can be used to discuss the precession and relaxation dynamics of the spin from another, namely from the electronic perspective.
This information is in principle experimentally accessible to spin-resolved scanning-tunnelling microscope techniques \cite{Wie09,NF93,LEL+10,Mor10} and important for an atomistic understanding of nano-spintronics devices. \cite{WAB01,KWCW11}
We are particularly interested in the physics of the system in the strong-$J$ regime or for a strong field $\ff B$ where the time scales of the spin and the electron dynamics become comparable. 
This has not yet been explored but could become relevant to understand real-time dynamics in realizations of strong-$J$ Kondo-lattice models by means of ultracold fermionic Yb quantum gases trapped in optical lattices. \cite{SHH+14,CMP+14}

The paper is organized as follows:
We first introduce the model and the equations of motion for the exact quantum-classical hybrid dynamics in Sec.\ \ref{sec:mt} and discuss some computational details in Sec.\ \ref{sec:num}.
Sec.\ \ref{sec:spindyn} provides a comprehensive discussion of the relaxation of the classical spin after a sudden switch of a magnetic field. 
The reversal time as a function of the interaction and the field strength is analyzed in detail. 
We then set the focus on the conduction-electron system which induces the relaxation of the classical spin by dissipation of energy.
In Sec.\ \ref{sec:spinonly}, the linear-response approach to integrate out the electron degrees of freedom is carefully examined, including a discussion of the additional approximations that are necessary to re-derive the LLG equation and the damping term in particular.
Sec.\ \ref{sec:con} summarizes the results and the main conclusions.

%--------------------------------------------------------------------------------------------------------------
\section{Model and theory}
\label{sec:mt}
%--------------------------------------------------------------------------------------------------------------

We consider a classical spin $\ff S$ with $|\ff S| = 1/2$, which is coupled via a local exchange interaction of strength $J$ to the local quantum spin $\ff s_{i_0}$ at the site $i_{0}$ of a system of $N$ itinerant and non-interacting conduction electrons.
The conduction electrons hop with amplitude $-T$ ($T>0$) between non-degenerate orbitals on nearest-neighboring sites of a $D$-dimensional lattice, see Fig.\ \ref{fig:system}.
$L$ is the number of lattice sites, and $n=N/L$ is the average conduction-electron density. 

The dynamics of this quantum-classical hybrid system \cite{Elz12} is determined by the Hamiltonian
\begin{equation}
{H} = - T \sum_{\langle ij \rangle, \sigma} c^{\dagger}_{i\sigma} c_{j\sigma} 
+ 
J \ff s_{i_0} \ff S 
-
\ff B \ff S 
\: .
\label{eq:ham}
\end{equation}
Here, $c_{i\sigma}$ annihilates an electron at site $i=1,...,L$ with spin projection $\sigma=\uparrow, \downarrow$, and 
$\ff s_{i} = \frac{1}{2} \sum_{\sigma \sigma'} c^{\dagger}_{i\sigma} \ff \sigma_{\sigma\sigma'} c_{i\sigma'}$ is the local conduction-electron spin at $i$, where $\ff \sigma$ denotes the vector of Pauli matrices.
The sum runs over the different ordered pairs $\langle i j \rangle$ of nearest neighbors. 
$\ff B$ is an external magnetic field which couples to the classical spin.

To be definite, an antiferromagnetic exchange coupling $J>0$ is assumed. 
If $\ff S$ was a quantum spin with $S=1/2$, Eq.\ (\ref{eq:ham}) would represent the single-impurity Kondo model. \cite{Kon64,Hew93}
However, in the case of a classical spin considered here, there is no Kondo effect.
The semiclassical single-impurity Kondo model thus applies to systems where a local spin is coupled to electronic degrees of freedom but where the Kondo effect absent or suppressed. 
This comprises the case of large spin quantum numbers $S$, or the case of temperatures well above the Kondo scale, or systems with a ferromagnetic Kondo coupling $J<0$ where, for a classical spin, we expect a qualitatively similar dynamics as for $J>0$.

We assume that initially, at time $t=0$, the classical spin $\ff S(t=0)$ has a certain direction and that the conduction-electron system is in the corresponding ground state, i.e., the conduction electrons occupy the lowest $N$ one-particle eigenstates of the non-interacting Hamiltonian Eq.\ (\ref{eq:ham}) for the given $\ff S=\ff S(t=0)$ up to the chemical potential $\mu$.
A non-trivial time evolution is initiated if the initial direction of the classical spin and the direction of the field $\ff B$ are non-collinear.

To determine the real-time dynamics of the electronic subsystem, it is convenient to introduce the reduced one-particle density matrix.
Its elements are defined as expectation values,
\begin{equation}
\rho_{ii',\sigma\sigma'}(t) \equiv \langle c_{i'\sigma'}^{\dagger} c_{i\sigma} \rangle_{t} \; ,
\label{eq:densitymatrix}
\end{equation}
in the system's state at time $t$.
At $t=0$ we have $\ff \rho(0) = \Theta (\mu - \ff T(0))$.
The elements of $\ff \rho(0)$ are given by 
\begin{equation}
  \rho_{i\sigma,i'\sigma'}(0) = \sum_{k} U_{i\sigma, k} \Theta (\mu - \varepsilon_{k}) U^{\dagger}_{k, i'\sigma'} \; , 
\end{equation}
where $\Theta$ is the step function and where $\ff U$ is the unitary matrix diagonalizing the hopping matrix $\ff T(0)$, i.e., $\ff U^{\dagger} \ff T(0) \ff U = \ff \varepsilon$ with the diagonal matrix $\ff \varepsilon$ given by the eigenvalues of $\ff T(0)$.
The hopping matrix at time $t$ is can be read off from Eq.\ (\ref{eq:ham}). 
It comprises the physical hopping and the contribution resulting from the coupling term. 
Its elements are given by
\begin{equation}
  T_{i\sigma,i'\sigma'}(t) = -T \delta_{\langle ii' \rangle} \delta_{\sigma\sigma'} + \delta_{ii_{0}} \delta_{i'i_{0}}
  \frac{J}{2} (\ff S(t) \ff \sigma)_{\sigma\sigma'} \: .
\label{eq:hopp}
\end{equation}
Here $\delta_{\langle ii' \rangle} =1$ if $i, i'$ are nearest neighbors and zero else.

There is a closed system of equations of motion for the classical spin vector $\ff S(t)$ and for the one-particle density matrix $\ff \rho(t)$. 
The time evolution of the classical spin is determined via $(d/dt) \ff S(t) = \{ \ff S , H^{\rm class.} \}$ by the classical Hamilton function $H^{\rm class.} = \langle H \rangle$.
This equation of motion is the only known way to consistently describe the dynamics of quantum-classical hybrids (see Refs.\ \onlinecite{Hes85,Hal08,Elz12} and references therein for a general discussion).
The Poisson bracket between arbitrary functions $A$ and $B$ of the spin components is given by, \cite{YH80,LD83}
\begin{equation}
  \{ A, B \} = \sum_{\alpha,\beta,\gamma} 
  \varepsilon_{\alpha\beta\gamma}
  \frac{\partial A}{\partial S_{\alpha}}
  \frac{\partial B}{\partial S_{\beta}}
  S_{\gamma} \; ,
\end{equation}
where the sums run over $x,y,z$ and
where $\varepsilon_{\alpha\beta\gamma}$ is the fully antisymmetric $\varepsilon$-tensor.
With this we find
\begin{equation}
\frac{d}{dt} \ff S(t)
=
J \langle \ff s_{i_0} \rangle_{t} \times \ff S(t)
-
\ff B \times \ff S(t) \: .
\label{eq:largecl}
\end{equation}
This is the Landau-Lifschitz equation where the expectation value of the conduction-electron spin at $i_{0}$ is given by
\begin{equation}
  \langle \ff s_{i_0} \rangle_{t} 
  =
  \frac{1}{2} \sum_{\sigma\sigma'} \rho_{i_{0} \sigma, i_{0}\sigma'}(t) \, \ff \sigma_{\sigma'\sigma} \: ,
\label{eq:si0}  
\end{equation}
and where $J \langle \ff s_{i_0} \rangle_{t}$ acts as an effective time-dependent internal field in addition to the external field $\ff B$.

The equation of motion for $\langle \ff s_{i} \rangle_{t}$ reads as
\begin{eqnarray}
\frac{d}{dt}\langle{\ff s_{i}} \rangle_{t} 
&=&
\delta_{ii_{0}} J\ff S(t) \times \langle \ff s_{i} \rangle_{t}
\nonumber \\
&+&
T
\sum_{j}^{n.n.}
\frac{1}{2i}
\sum_{\sigma\sigma'}
(\langle c^{\dagger}_{i\sigma} \ff \sigma_{\sigma\sigma'} c_{j\sigma'}
\rangle_{t} - \mbox{c.c.}) \: ,
\label{eq:smallcl}
\end{eqnarray}
where the sum runs over the nearest neighbors of $i$.
The second term on the right-hand side describes the coupling of the local conduction-electron spin to its environment and the dissipation of spin and energy into the bulk of the system (see below).
Apparently, the system of equations of motion can only be closed by considering the complete one-particle density matrix Eq.\ (\ref{eq:densitymatrix}).
It obeys a von Neumann equation of motion,
\begin{equation}
  i \frac{d}{dt} \ff \rho(t) = [ \ff T(t) , \ff \rho(t) ] \: 
\label{eq:rhoeom}
\end{equation}
as is easily derived, e.g., from the Heisenberg equation of motion for the annihilators and creators.

As is obvious from the equations of motion, the real-time dynamics of the quantum-classical Kondo-impurity model on a lattice with a finite but large number of sites $L$ can be treated numerically exactly (see also below).
Nevertheless, the model comprises highly non-trivial physics as the electron dynamics becomes effectively correlated due to the interaction with the classical spin. 
In addition, the effective electron-electron interaction mediated by the classical spin is retarded: 
electrons scattered from the spin at time $t$ will experience the effects of the spin torque exerted by electrons that have been scattered from the spin at earlier times $t'<t$.

%--------------------------------------------------------------------------------------------------------------
\section{Computational details}
\label{sec:num}
%--------------------------------------------------------------------------------------------------------------

%--------------------------------------------------------------------------------------------------------------
\begin{figure}[t]
\centering
\includegraphics[width=0.5\columnwidth]{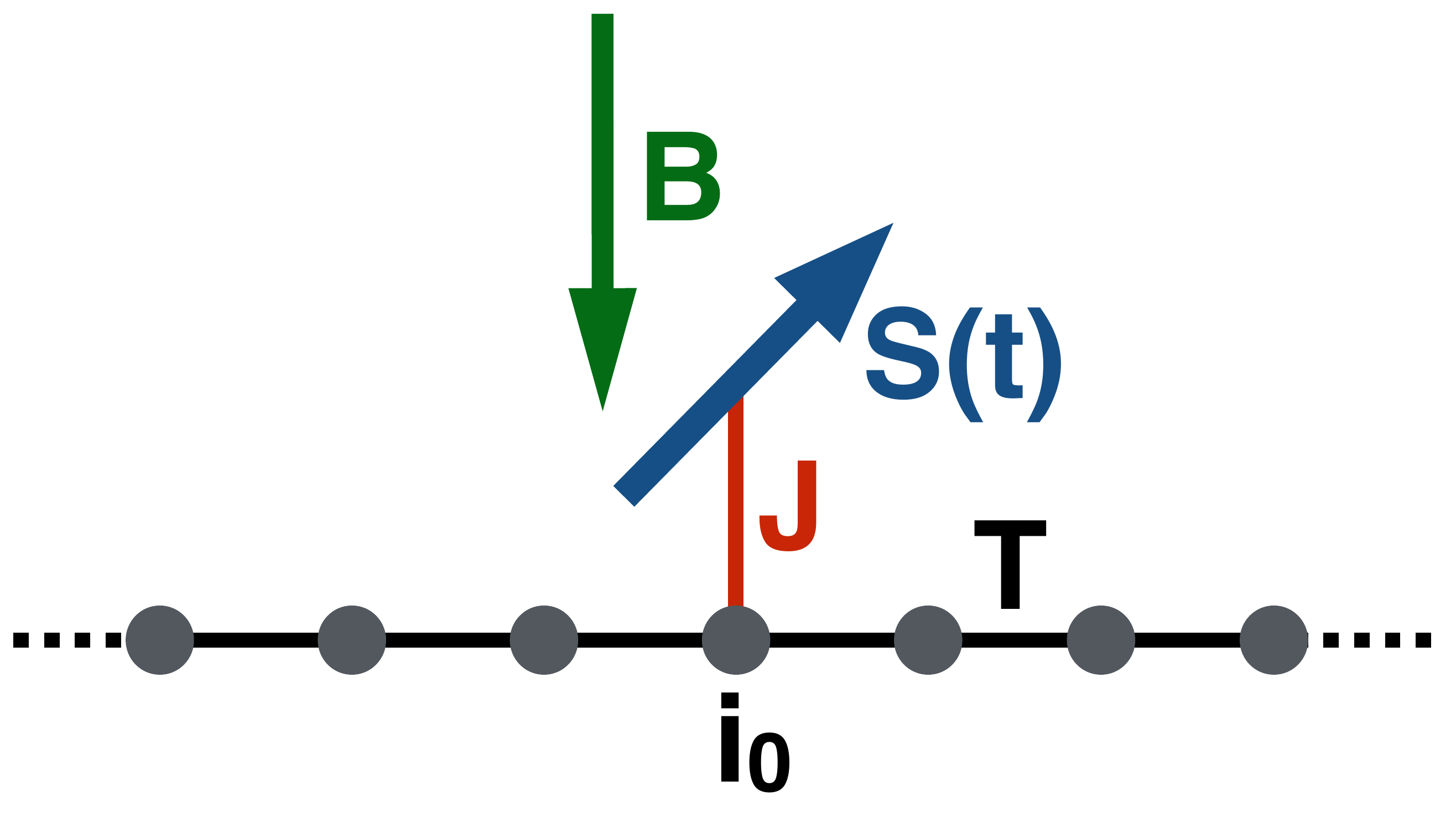}
\caption{(Color online)
Classical spin $\ff S(t)$ coupled via an antiferromagnetic local exchange interaction of strength $J$ to a system of conduction electrons hopping with nearest-neighbor hopping amplitude $T$ over the sites of a one-dimensional lattice with open boundaries.
The spin couples to the central site $i_{0}$ of the system and is subjected to a local magnetic field of strength $\ff B$.
} 
\label{fig:system}
\end{figure}
%--------------------------------------------------------------------------------------------------------------

Eqs.\ (\ref{eq:hopp}), (\ref{eq:largecl}), (\ref{eq:si0}) and (\ref{eq:rhoeom}) represent a coupled non-linear system of first-order ordinary differential equations 
which can be solved numerically. 
By blocking up the von Neumann equation (\ref{eq:rhoeom}), the differential equations are written in a standard form $\dot{\ff y}=\ff f(\ff y(t),t)$, where $\ff y(t)$ is a high-dimensional vector, such that an explicit Runge-Kutta method can be applied. 
A high-order propagation technique is used \cite{JHV2010} which provides the numerically exact solution up to 6-th order in the time step $\Delta t$. 
For a typical system consisting of about $L=10^{3}$ sites this implies that $\sim 10^{6}$ coupled equations are solved.

We consider a one-dimensional system with open boundaries consisting of $L=1001$ sites and a local perturbation at the central site $i_{0}$ of the system, see Fig.\ \ref{fig:system}.
For a half-filled tight-binding conduction band the Fermi velocity $v_{\rm F} = 2T$ roughly determines the maximum speed of the excitations and defines a ``light cone''. \cite{LR72,BHV06}
This means that finite-size effects due to scattering at the system boundaries become relevant after a propagation time $t_{\rm max} \sim 500$ (in units of $1/T$).
A time step $\Delta t = 0.1$ is usually sufficient for reliable numerical results up to $t_{\rm max}$, i.e., about 5000 time steps are performed. 
The computational cost is moderate, and calculations can be performed in a few hours on a standard desktop computer.

Assuming, for example, that $\ff B =  (0,0,B)$, the Hamiltonian is invariant under rotations around the $z$ axis.
It is then easily verified that not only the length of the spin $|\ff S|=1/2$ is conserved but also the total number of conduction electrons,
\begin{equation}
N_{\rm tot} = \sum_{i\sigma} \langle c^{\dagger}_{i\sigma} c_{i\sigma} \rangle \: , 
\label{eq:consn}
\end{equation}
the $z$-component of the total spin,
\begin{equation}
S_{\rm tot, z} = S_{z} + \sum_{i} \langle s_{iz} \rangle \: ,
\label{eq:conss}
\end{equation}
as well as the total energy,
\begin{equation}
E_{\rm tot} = \langle H \rangle = \tr (\ff \rho(t) \ff T(t)) - \ff B \ff S(t) \: .
\label{eq:conse}
\end{equation}
Despite the fact that the model does not include a direct (e.g., Coulomb) interaction among the conduction electrons, the average occupation numbers of the basis of one-particle states in which the hopping matrix $\ff T(t)$ is diagonal at time $t$ are {\em not} conserved. 
This is due to the effective retarded interaction mediated by the classical spin.
Hence, the system is not integrable, unlike a free fermion gas.
The conservation of the above-mentioned global observables serves as a sensitive check for the accuracy of the numerical procedure.

%--------------------------------------------------------------------------------------------------------------
\section{Coupled spin and electron dynamics}
\label{sec:spindyn}
%--------------------------------------------------------------------------------------------------------------

%--------------------------------------------------------------------------------------------------------------
\subsection{Spin relaxation}
\label{sec:rel}
%--------------------------------------------------------------------------------------------------------------

Fig.\ \ref{fig:bloch} shows the real-time dynamics of the classical spin for $J=1$. 
Energy and time units are fixed by the nearest-neighbor hopping $T=1$ throughout the paper. 
Initially, for $t=0$, the spin is oriented (almost) {\em antiparallel} to the external local field $\ff B = (0,0,B)$ with $B=1$, i.e., initially
$S_x(0) = \frac{1}{2} \sin\vartheta$, $S_{y}(0) = 0$, $S_z(0) = - \frac{1}{2} \cos\vartheta$ where a non-zero but small polar angle $\vartheta = \pi/50$ is necessary to slightly break the symmetry of the initial state and to start the dynamics.

For the same setup, the Landau-Lifschitz-Gilbert equation would essentially predict two effects:
first, a precession of the classical spin around the field direction with Larmor frequency $\omega_{L} = B_{z}$, and second, a relaxation of the spin to the equilibrium state with $\ff S$ {\em parallel} to $\ff B = B \ff e_{z}$ for $t\to \infty$.
Both effects are also found in the full dynamics of the quantum-classical hybrid model. 
The frequency of the oscillation of $S_{x}(t)$ that is seen in Fig.\ \ref{fig:bloch} is $\omega_{L}$, and $S_{z}$ is reversed after a few hundred time units. 
The precessional motion is easily explained by the torque on the spin exerted by the field according to Eq.\ (\ref{eq:largecl}). 
The explanation of the damping effect is more involved:

%--------------------------------------------------------------------------------------------------------------
\begin{figure}[t]
\centering
\includegraphics[width=\columnwidth]{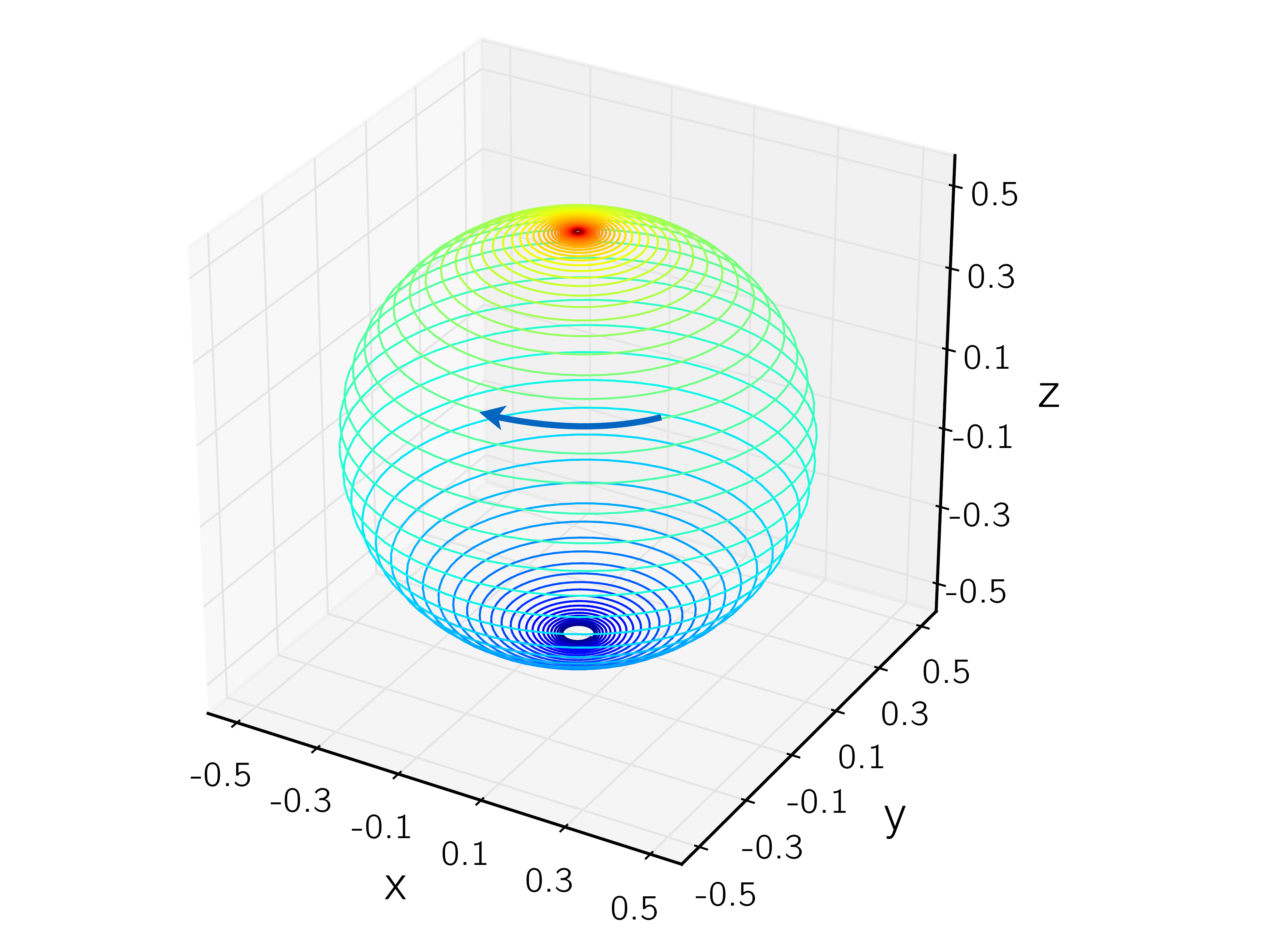}
\includegraphics[width=0.94\columnwidth]{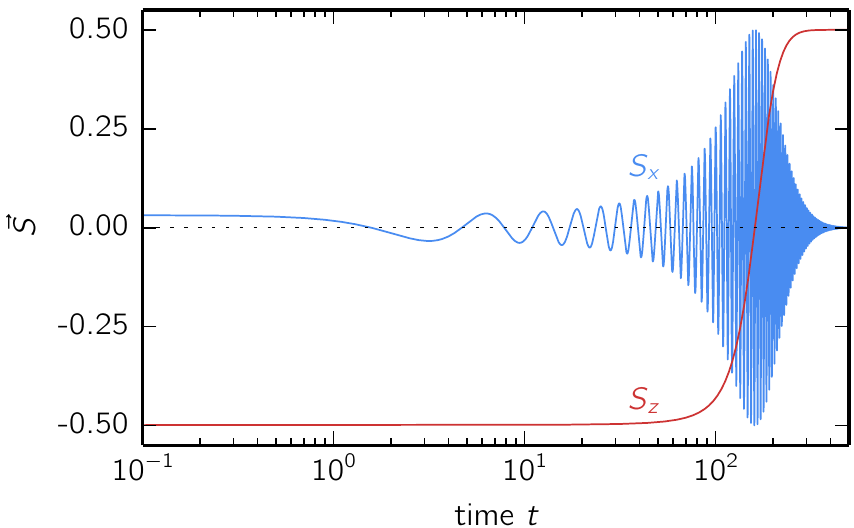}
\caption{(Color online)
Real-time dynamics of the classical spin.
{\em Upper panel:} Bloch sphere representation.
{\em Lower panel:} $x$ and $z$ component of $\ff S(t)$ ($|\ff S|=1/2$).
Calculations for exchange coupling $J=1$ and field strength $B=1$ and for a system of $L=1001$ sites.
($L=1001$ is kept fixed for the rest of the paper).
Energy and time units are fixed by the nearest-neighbor hopping $T=1$. 
} 
\label{fig:bloch}
\end{figure}
%--------------------------------------------------------------------------------------------------------------

%--------------------------------------------------------------------------------------------------------------
\begin{figure}[t]
\centering
\includegraphics[width=\columnwidth]{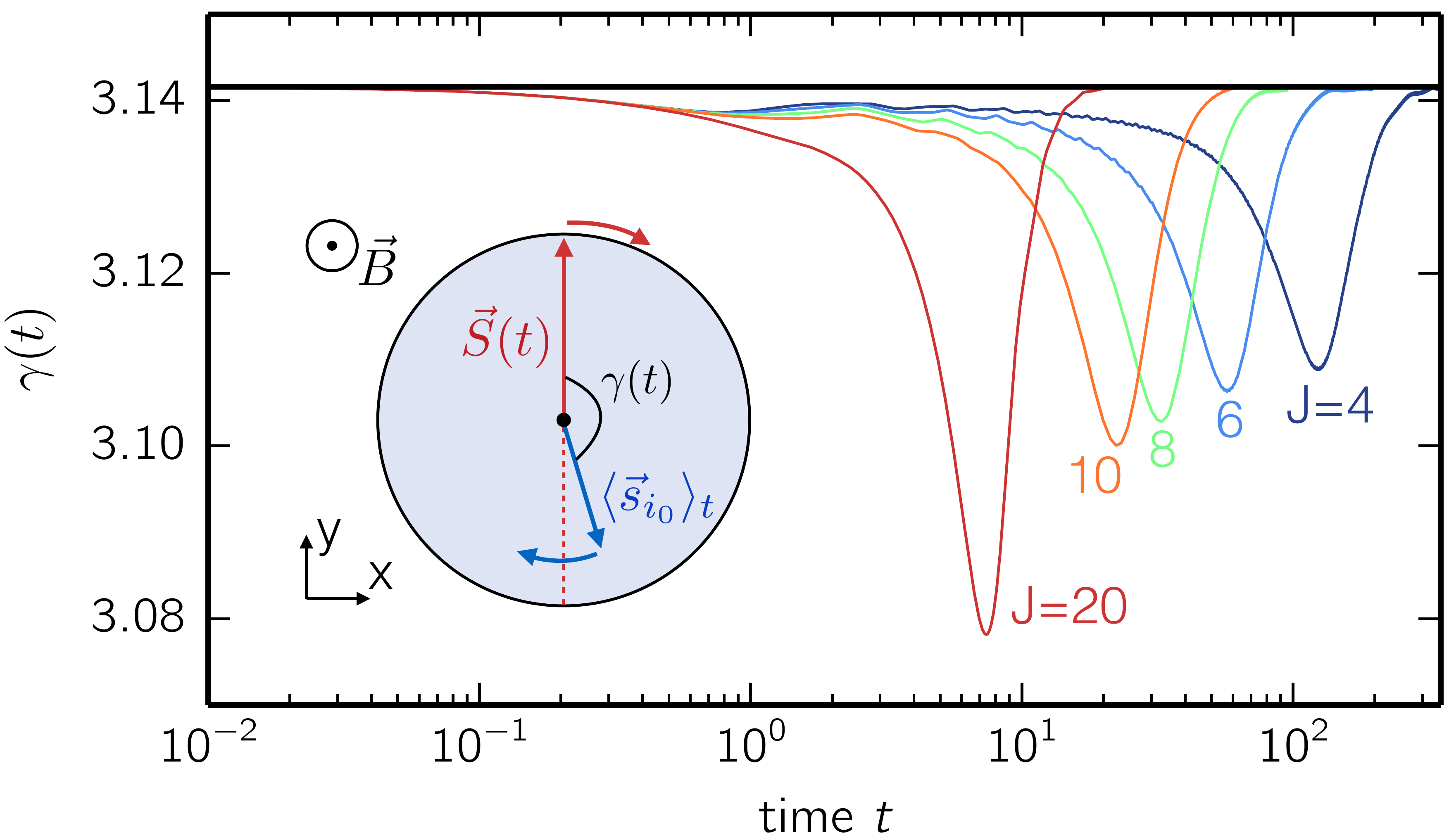}
\caption{(Color online)
Time dependence of the angle $\gamma(t)$ enclosed by $\ff S(t)$ and $\langle \ff s_{i_{0}} \rangle_{t}$ for $B=0.1$ and different $J$ as indicated. 
} 
\label{fig:angle}
\end{figure}
%--------------------------------------------------------------------------------------------------------------

Even for the high field strength considered here, the spin dynamics is slow as compared to the characteristic electronic time scale such that it could be reasonable to assume the electronic system being in its instantaneous ground state at any instant of time and corresponding to the configuration of the classical spin. 
This, however, would imply that the expectation value $\langle \ff s_{i_{0}} \rangle_{t}$ of the local conduction-electron moment at $i_{0}$ is always strictly parallel to $\ff S(t)$ and, hence, there would not be any torque mediated by the exchange coupling $J$ on $\ff S(t)$.

In fact, the direction of $\langle \ff s_{i_{0}} \rangle_{t}$ is always somewhat {\em behind} the ``adiabatic direction'', i.e., behind $-\ff S(t)$:
This is shown in Fig.\ \ref{fig:angle} for a field of strength $B=0.1$ where the spin dynamics is by a factor 10 slower, compared to Fig.\ \ref{fig:bloch}, and for different stronger exchange couplings $J$.
Even in this case the process is by no means adiabatic, and the angle $\gamma(t)$ between $\ff S(t)$ and $\langle \ff s_{i_{0}} \rangle_{t}$ is close to but clearly smaller than $\gamma=\pi$ at any instant of time.
This non-adiabaticity results from the fact that the motion of the classical spin affects the conduction electrons in a retarded way, i.e., it takes a finite time until the local conduction-electron spin $\langle \ff s_{i_{0}} \rangle_{t}$ at $i_{0}$ reacts to the motion of the classical spin.

This retardation effect results in a torque $J \langle \ff s_{i_0} \rangle_{t} \times \ff S(t) \ne 0$ exerted on the classical $\ff S(t)$ in the $+z$ direction which adds to the torque due to $\ff B$ and which drives the spin to its new equilibrium direction. 
Hence, retardation is the physical origin of the Gilbert spin damping.

With increasing time, the deviation of $\gamma(t)$ from the instantaneous equilibrium value $\gamma=\pi$ increases in magnitude, i.e., the direction of $\langle \ff s_{i_{0}} \rangle_{t}$ is more and more behind the adiabatic direction, and the torque increases. 
Its magnitude is at a maximum at the same time when the oscillating $x, y$ components of $\ff S(t)$ are at a maximum (see Fig.\ \ref{fig:bloch}). 
The $z$-component of the torque does not vanish before the spin has reached its new equilibrium position $\ff S(t) \propto \ff e_{z}$.

Fig.\ \ref{fig:angle} shows results for different $J$. 
Generally, non-adiabatic effects show up if the typical time scale of the dynamics is faster than the relaxation time, i.e., the time necessary to transport the excitation away from the location $i_{0}$ where it is created initially.
Roughly this time scale is set by the inverse hopping $1/T$.
One therefore expects that, for fixed $T$, a stronger $J$ implies a stronger retardation of the conduction-electron dynamics.
The results for different $J$ shown in Fig.\ \ref{fig:angle} in fact show that the maximum deviation of $\gamma(t)$ from the adiabatic direction $\gamma=\pi$ increases with increasing $J$ (for very strong $J$ the dynamics becomes much more complicated, see below).
This results in a stronger torque on $\ff S(t)$ in $z$ direction and thus in a stronger damping.
The picture is also qualitatively consistent with the LLG equation as the Gilbert damping constant $\alpha$ increases with $J$.

\begin{figure}[t]
\centering
\includegraphics[width=0.94\columnwidth]{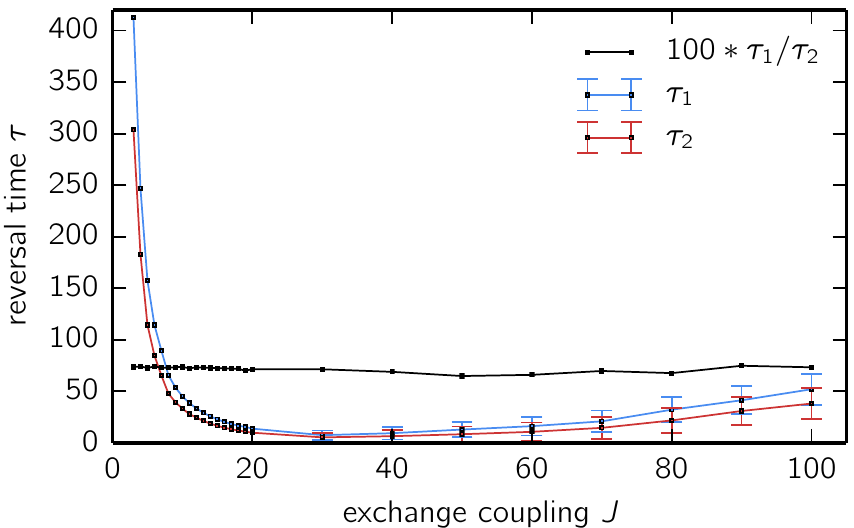}
\caption{(Color online)
Time for a spin reversal $S_z = \frac{1}{2} \cos(\pi - \vartheta)  = - \frac{1}{2} \cos(\vartheta) \to S_{z} = \frac{1}{2} \cos(\vartheta)$ 
for $\vartheta = \vartheta_{1} = \pi/50$ (reversal time $\tau_{1}$) and 
for $\vartheta = \vartheta_{2} = \pi/25$ (reversal time $\tau_{2}$).
Calculations as a function of $J$ for fixed $B=0.1$. 
} 
\label{fig:rev}
\end{figure}

The spin (almost) reverses its direction after a finite reversal time $\tau$ which is shown in Fig.\ \ref{fig:rev} as a function of $J$. 
Calculations have been performed for an initial direction of the classical spin with $S_{z}(0) = - (1/2) \cos \vartheta$ with two different polar angles $\vartheta_{1,2}$. 
If $\vartheta$ is sufficiently small, the results for different $\vartheta$ are expected to differ by a $J$ and $B$ independent constant factor only. 
As Fig.\ \ref{fig:rev} demonstrates, the ratio $\tau_{1}/\tau_{2}$ is in fact nearly constant.
For weak $J$ and up to coupling strengths of about $J\lesssim 30$, we find that the reversal time decreases with increasing $J$. 
With increasing $J$, the retardation effect increases, as discussed above, and the stronger damping results in a shorter reversal time. 

%--------------------------------------------------------------------------------------------------------------
%--------------------------------------------------------------------------------------------------------------

%--------------------------------------------------------------------------------------------------------------
\begin{figure}[b]
\centering
\includegraphics[width=0.94\columnwidth]{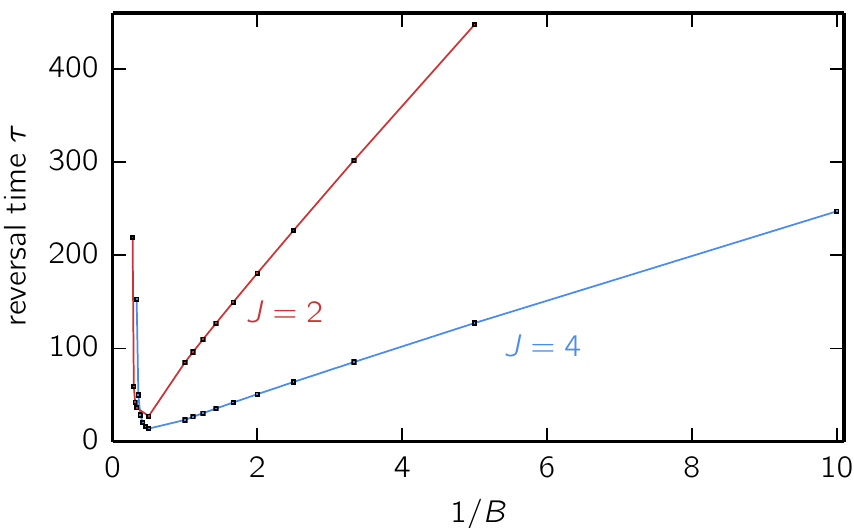}
\caption{(Color online)
Reversal time $\tau_{1}$ as function of $B$ for fixed $J=2$ and $J=4$ as indicated.
} 
\label{fig:rev2}
\end{figure}
%--------------------------------------------------------------------------------------------------------------

The prediction of the LLG equation for the reversal time of a single spin $\tau$ can be derived analytically \cite{Kik56} and is given by
\begin{equation}
  \tau \propto \frac{1+\alpha^{2}}{\alpha} \frac{1}{B} \ln \Big| \frac{1/2-S_{z}(0)}{1/2+S_{z}(0)} \Big| \: .
\label{eq:rtime}
\end{equation}
However, down to the smallest $J$ for which $\tau$ can be calculated reliably, our results for the full spin dynamics do not scale as $\tau \propto 1/J^{2}$ as one would expect for weak $J$ assuming that $\alpha \propto J^{2}$ (see discussion in Sec.\ \ref{sec:llg}). 

The $B$ dependence of the reversal time is shown in Fig.\ \ref{fig:rev2}. 
With increasing field strength, the classical spin $\ff S(t)$ precesses with a higher Larmor frequency $\omega_{L} \approx B$ around the $z$ axis.
Hence, non-adiabatic effects increase.
The stronger the field, the more delayed is the precessional motion of the local conduction-electrons spin $\langle \ff s_{i_{0}} \rangle_{t}$. 
This results in a stronger torque in $+z$ direction exerted on the classical spin. 
Therefore, the relaxation is faster and the reversal time $\tau$ smaller. 
For weak and intermediate field strengths, $\tau$ is roughly proportional to $1/B$. 
This is consistent with the prediction of the LLG equation, see Eq.\ (\ref{eq:rtime}). 

In the limit of very strong fields one would expect an increase of the reversal time with increasing $B$ since the field term will eventually dominate the dynamics, i.e., only the precessional motion survives which implies a diverging reversal time. 
In fact, for a field strength exceeding a critical strength $B_{c}$, which depends on $J$, there is no full relaxation any longer, and $\tau = \infty$. 
This strong-$B$ regime cannot be captured by the LLG equation and deserves further studies.

%--------------------------------------------------------------------------------------------------------------
\begin{figure}[t]
\centering
\includegraphics[width=\columnwidth]{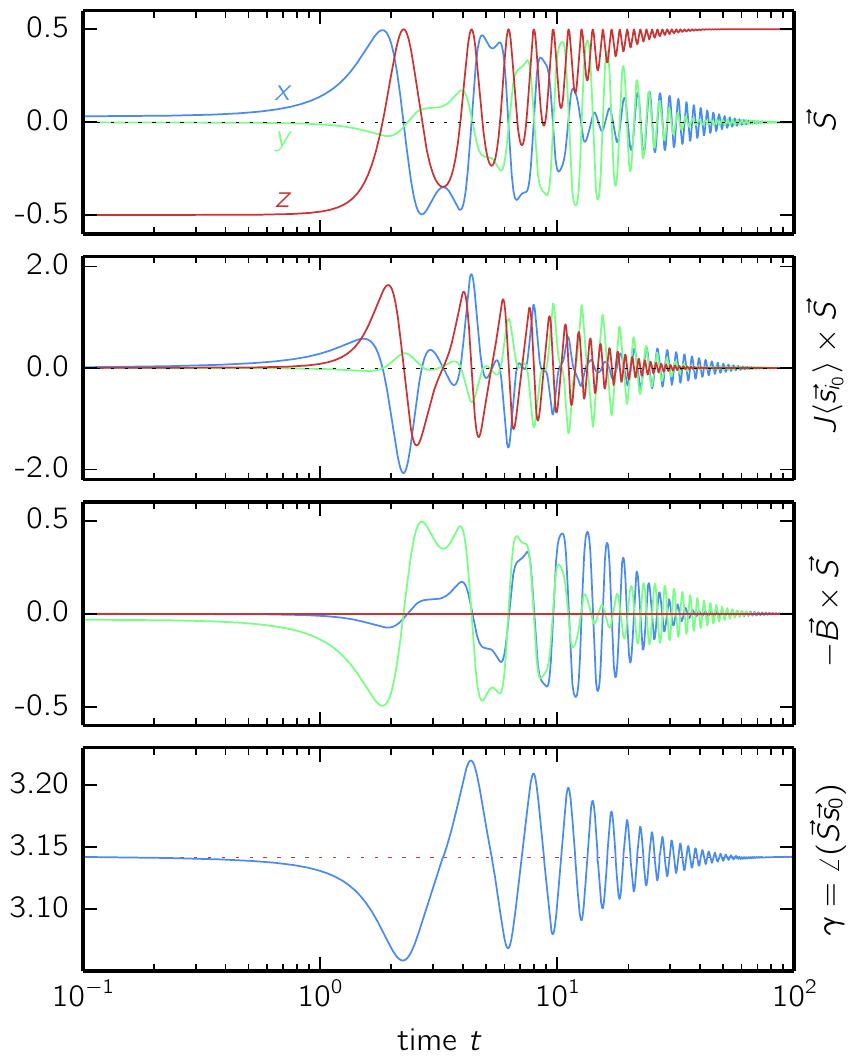}
\caption{(Color online)
Real-time dynamics in the strong-$J$ regime: $J=100$ and $B=0.1$.
{\em First panel:} Classical spin $\ff S(t)$. 
{\em Second panel:} Torque on $\ff S(t)$ due to the exchange interaction.
{\em Third panel:} Torque on $\ff S(t)$ due to the field term. 
{\em Fourth panel:} Angle enclosed by 
$\ff S(t)$ and $\langle \ff s_{i_{0}} \rangle_{t}$
} 
\label{fig:sb1}
\end{figure}
%--------------------------------------------------------------------------------------------------------------

The strong-$J$ regime is interesting as well. 
For coupling strengths exceeding $J \approx 30$ the reversal time {\em increases} with $J$ (see Fig.\ \ref{fig:rev}).
Eventually, the reversal time must even diverge. 
This is obvious as the dynamics is described by a simple two-spin model in the limit $J = \infty$ which cannot show spin relaxation.
The corresponding equations of motion are obtained by from Eqs.\ (\ref{eq:largecl}) and (\ref{eq:smallcl}) by setting $t=0$ and $\ff B=0$:
\begin{eqnarray}
\frac{d}{dt} \ff S(t)
&=&
J \langle \ff s_{i_0} \rangle_{t} \times \ff S(t) \: ,
\nonumber \\
\frac{d}{dt}\langle{\ff s_{i_{0}}} \rangle_{t} 
&=&
J\ff S(t) \times \langle \ff s_{i_{0}} \rangle_{t}
\end{eqnarray}
Note that we have $|\langle \ff s_{i_{0}} \rangle_{t}|=1/2$ for $J\to \infty$.
The equations are easily solved by exploiting the conservation of the total spin $\ff S_{\rm tot} = \ff S(t) + \langle \ff s_{i_{0}} \rangle_{t}$. 
Both spins precess with constant frequency $\omega_{0} = J S_{\rm tot}$ around $\ff S_{\rm tot}$. 
Their components parallel to $\ff S_{\rm tot}$ are equal, and their components perpendicular to $\ff S_{\rm tot}$ are anti-parallel and of equal length. 

However, the two-spin dynamics of the $J = \infty$ limit is not stable against small perturbations. 
Fig.\ \ref{fig:sb1} shows the classical spin dynamics of the full model (with $T=1$ and $B=0.1$) for a very strong but finite coupling $J=100$. 
Here, the motion of the classical spin gets very complicated as compared with the highly regular behavior in the weak-$J$ regime (cf.\ Fig.\ \ref{fig:bloch}).
In particular, the $z$-component of $\ff S(t)$ is oscillating on nearly the same scale as the $x$ and $y$ components. 
This characteristic time scale $\Delta t \approx 4$ of the oscillation corresponds to a frequency $\omega \approx 1.5$ which differs by more than an order of magnitude from both, the Larmor frequency $B = 0.1$ and from the exchange-coupling strength $J=100$. 
Note that the oscillation of $S_{z}(t)$ is actually the reason for the ambiguity in the determination of the precise reversal time and gives rise to the error bars in Fig.\ \ref{fig:rev} for strong $J$. 

We attribute the complexity of the dynamics to the fact that the torque due to the field term and the torque due to the exchange coupling are of comparable magnitudes, see second and third panel in Fig.\ \ref{fig:sb1}. 
It is interesting that even for strong $J$, where one would expect $\ff S(t)$ and $\langle \ff s_{i_{0}} \rangle_{t}$ to form a tightly bound local spin-zero state, there is actually a small deviation from perfect antiparallel alignment, i.e., $\gamma(t) \ne \pi$, as can be seen in the fourth panel of Fig.\ \ref{fig:sb1}. 
This results in a finite $z$-component of the torque on $\ff S(t)$ which leads to a very fast reversal with $S_{z}(t) \approx +0.5$ at time $t\approx 2.1$.
Contrary to the weak-coupling limit, however, the $z$-component of the torque changes sign at this point and drives the spin back to the $-z$-direction.
At $t \approx 3.2$, however, the $z$-component of $\ff S(t)$ once more reverses its direction. 
Here, the torque due to the exchange coupling vanishes completely as $\ff S(t)$ and $\langle \ff s_{i_{0}} \rangle_{t}$ are perfectly antiparallel (see the first zero of $\gamma(t)$ in the fourth panel).
The motion continues due to the non-zero field-induced torque. 
This pattern repeats several times. 
$\ff S(t)$ mainly oscillates within a plane including and slowly rotating around the $z$-axis.

Eventually, there is a perfect relaxation of the classical spin for large $t$ but in a very different way as compared to the weak-coupling limit. 
While the deviation from $\gamma = \pi$ is small at any instant of time as for weak $J$, the most apparent difference is perhaps that the new ground state is approached with an oscillating behavior of $\gamma(t)$ around $\gamma=\pi$, i.e., $\langle \ff s_{i_{0}} \rangle_{t}$ may run behind or {\em ahead of} $\ff S(t)$ as well.  

Let us summarize at this point the main differences between the quantum-classical hybrid and the effective LLG dynamics of the classical spin:
For weak $J$ and $B$, the qualitative behavior, precessional motion and relaxation, is the same in both approaches.
Quantitatively, however, the LLG equation is inconsistent with the observed $J$ dependence of the reversal time when assuming $\alpha \propto J^{2}$. 
The $B$ dependence of $1/\tau$ is linear as expected from the LLG approach.
For strong $J$, the spin dynamics qualitatively differs from LLG dynamics and gets more complicated with a new time scale emerging.
Absence of complete relaxation, as observed in the strong-$B$ limit, is also not accessible to an effective spin-only theory. 

%--------------------------------------------------------------------------------------------------------------
\subsection{Energy dissipation}
\label{sec:edis}
%--------------------------------------------------------------------------------------------------------------

To complete the picture of the relaxation dynamics of the classical spin, the discussion should also comprise the dynamics of the electronic degrees of freedom. 
The spin relaxation must be accompanied by a dissipation of energy and spin into the bulk of the electronic system since the total energy and the total spin are conserved quantities, see Eqs.\ (\ref{eq:conss}) and (\ref{eq:conse}), while 
conservation of the total particle number, Eq.\ (\ref{eq:consn}), is trivially ensured by the particle-hole symmetric setup considered here where the average conduction-electron number at every site is time-independent: $\sum_{\sigma} \langle n_{i\sigma} \rangle_{t} = 1$. 

%--------------------------------------------------------------------------------------------------------------
\begin{figure}[t]
\centering
\includegraphics[width=\columnwidth]{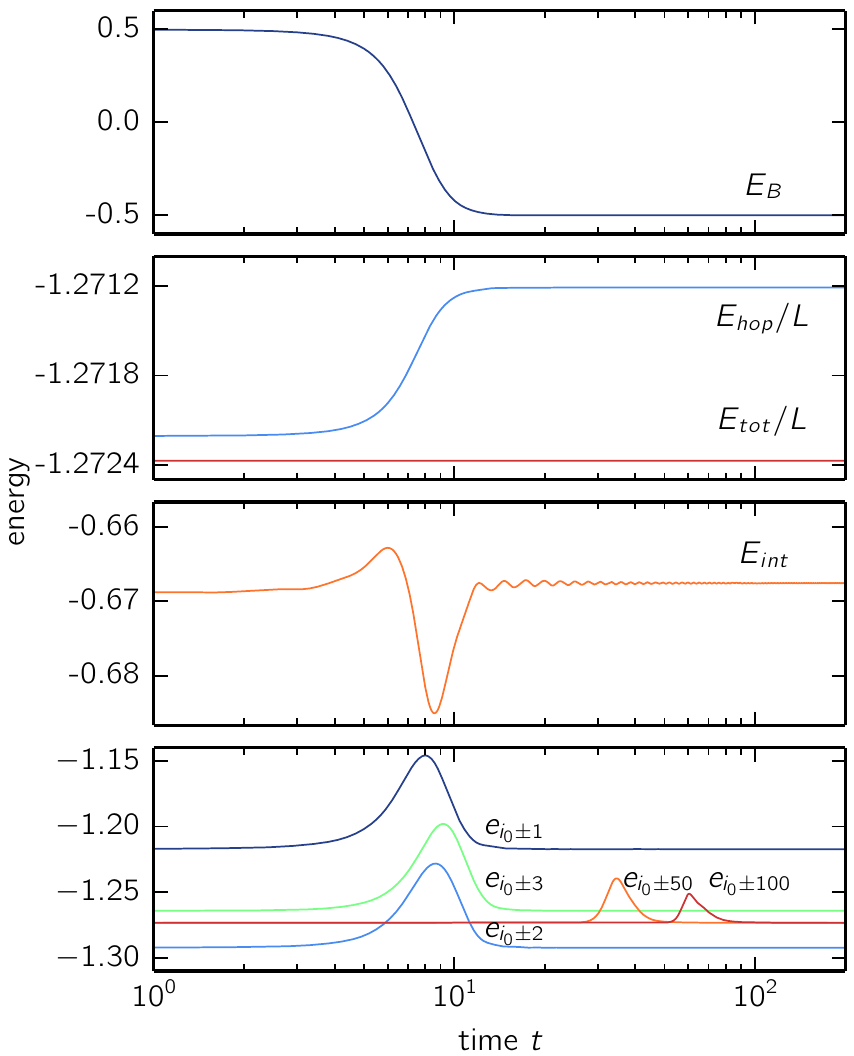}
\caption{(Color online)
Different contributions to the total energy as functions of time for $J=5$ and $B=1$.
{\em Top panel}: $E_{\rm B}$, energy of the classical spin in the external field. 
{\em Second panel}: $E_{\rm hop}/L$, kinetic (hopping) energy of the conduction-electron system per site, and $E_{\rm tot}/L$, total energy per site.
{\em Third panel}: $E_{\rm int}$, exchange-interaction energy.
The fourth panel shows the time-dependence of the total-energy density at different distances from the site $i_{0}$.
} 
\label{fig:econt}
\end{figure}
%--------------------------------------------------------------------------------------------------------------

%--------------------------------------------------------------------------------------------------------------
\begin{figure}[t]
\centering
\includegraphics[width=0.97\columnwidth]{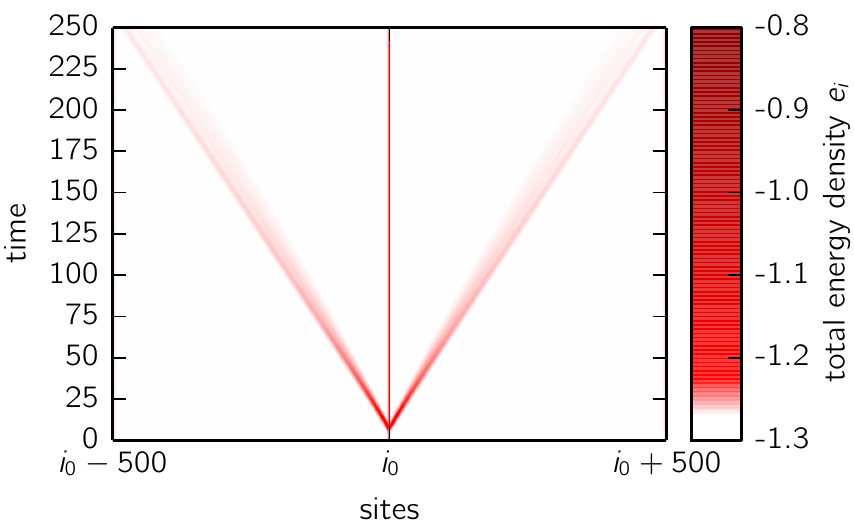}
\caption{(Color online)
Spatiotemporal evolution of the total-energy density $e_{i}(t)$, as defined in Eq.\ (\ref{eq:totedens}). 
Calculation for $J=5$, $B=1$. 
} 
\label{fig:edens}
\end{figure}
%--------------------------------------------------------------------------------------------------------------

The total energy is given by $E_{\rm tot} = \langle H \rangle$, see Eq.\ (\ref{eq:conse}), and is a sum over different contributions, 
$E_{\rm tot} =  E_{\rm B}(t) + E_{\rm hop}(t) + E_{\rm int}(t)$, namely the
interaction energy with the field
\begin{equation}
E_{\rm B}(t)
= 
- \ff B \ff S(t) \: , 
\end{equation}
the kinetic (hopping) energy of the conduction-electron system
\begin{equation}
  E_{\rm hop}(t)
  =
  - T 
  \sum_{\langle i j \rangle} \sum_{\sigma} \langle c^{\dagger}_{i\sigma} c_{j\sigma} \rangle_{t} \; ,
\end{equation}
and the exchange-interaction energy
\begin{equation}
  E_{\rm int}(t)
  =
  J 
  \langle \ff s_{i_0} \rangle_{t} \ff S(t) \: .
\end{equation}
The time dependence of those contributions is shown in Fig.\ \ref{fig:econt} for $J=5$ and $B=1$.

The top panel of Fig.\ \ref{fig:econt} shows that the system releases the interaction energy $|2\ff B \ff S|$ of the classical spin in the external field by aligning the spin to the field direction.
In the long-time limit, this energy is stored in the conduction-electron system: 
The average kinetic energy per site ($L=1001$) increases by the same amount as shown in the second panel (note the different scales).
The exchange-interaction energy changes with time but is the same for $t=0$ and $t\to \infty$ (see third panel).
The total energy is constant (second panel).

The relaxation of the classical spin in the external field $\ff B$ implies an energy flow away from the site $i_{0}$ into the bulk of the conduction-electron system such that {\em locally}, in the vicinity of $i_{0}$ the system is in its new ground state.
To discuss this energy flow, it is convenient to consider the total energy as a lattice sum $E_{\rm tot} = \sum_{i} e_{i}(t)$ over the total energy ``density'' defined as
\begin{equation}
e_{i}(t)
= 
- T \sum_{j}^{{\rm n.n.} (i)} \sum_{\sigma} \langle c^{\dagger}_{i\sigma} c_{j\sigma} \rangle_{t}
+ 
\delta_{ii_{0}}
(J \langle \ff s_{i_0} \rangle_{t} - \ff B) \ff S(t) 
\: ,
\label{eq:totedens}
\end{equation}
where the sum over $j$ runs over the nearest neighbors of site $i$.
The time dependence of the energy density in the vicinity of $i_{0}$ and at distances 50 and 100 is shown in the fourth panel of Fig.\ \ref{fig:econt}.

At any site in the conduction-electron system, the energy density increases from its ground-state value, reaches a maximum and eventually relaxes to the energy density of the new ground state. 
Since the latter is just the ground state with the reversed classical spin, the new ground-state energy density is the same as in the initial state at $t=0$. 
As can be seen in the fourth panel of Fig.\ \ref{fig:econt}, there is also a slight spatial oscillation of the ground-state energy density which just reflects the Friedel oscillations around the impurity at $i_{0}$.

Complete relaxation means that the excitation energy is completely removed from the vicinity of $i_{0}$ and transported into the bulk of the system. 
That this is in fact the case can be seen by comparing the energy density at different distances from $i_{0}$. 
It is also demonstrated by Fig.\ \ref{fig:edens} which visualizes the energy-current density which symmetrically points away from $i_{0}$:
The total energy of the excitation flowing through each pair of sites $i_{0} \pm \Delta i$ is constant, i.e., the time-integrated energy flux, $\int dt \,e_{i}(t)$ is the same for all $i$.

As is seen in Fig.\ \ref{fig:econt} (fourth panel) and Fig.\ \ref{fig:edens}, there is a considerable dispersion of the excitation wave packet carrying the energy. 
For example, at $i_{0}+100$ it takes more than four times longer, as compared to $i_{0}+1$, until most of the excitation has passed through (note the logarithmic time scale).

The broadening of the wave packet, due to dispersion, is asymmetric and bound by an upper limit for the speed of the excitation which is roughly set by the Fermi velocity $v_{\rm F} = 2T$.
This Lieb-Robinson bound \cite{LR72,BHV06} determines the ``light cone'' seen in Fig.\ \ref{fig:edens}.

%--------------------------------------------------------------------------------------------------------------
\subsection{Spin dissipation}
\label{sec:sdis}
%--------------------------------------------------------------------------------------------------------------

\begin{figure}[t]
\centering
\includegraphics[width=0.97\columnwidth]{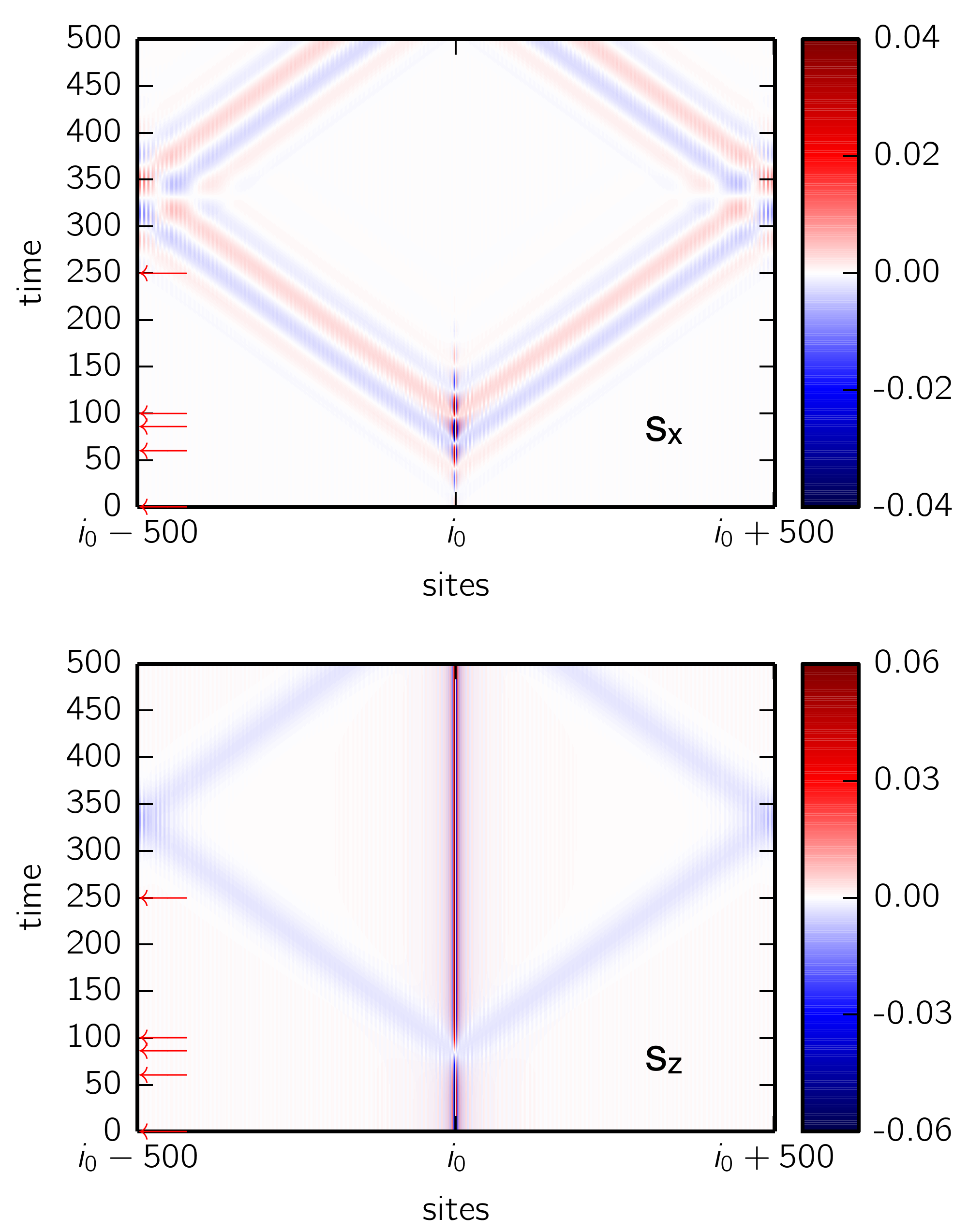}
\caption{(Color online)
Spatiotemporal evolution of the total-spin density $\langle \ff s_{i} \rangle_{t}$ 
(upper panel: $x$-component, lower panel $z$-component) for $J=5$ and $B=0.1$.
See Fig.\ \ref{fig:cut} for snapshots at times indicated by the arrows.
} 
\label{fig:spinden}
\end{figure}

The same upper speed limit, given by the Fermi velocity of the conduction-electron system, is also seen in the spatiotemporal evolution of the conduction-electron spin density $\langle \ff s_{i} \rangle_{t}$.
This is shown in Fig.\ \ref{fig:spinden} for a different magnetic field strength $B=0.1$ where the classical spin dynamics is slower.
Apparently, the wave packet of excitations emitted from the impurity not only carries energy but also spin. 
It symmetrically propagates away from $i_{0}$ and, at $t\approx 300$, reaches the system boundary where it is reflected perfectly. 
Up to $t=500$ there is hardly any effect visible in the local observables close to $i_{0}$ that is affected by the finite system size.

%--------------------------------------------------------------------------------------------------------------
%--------------------------------------------------------------------------------------------------------------

%--------------------------------------------------------------------------------------------------------------
\begin{figure*}[t]
\centering
\includegraphics[width=1.7\columnwidth]{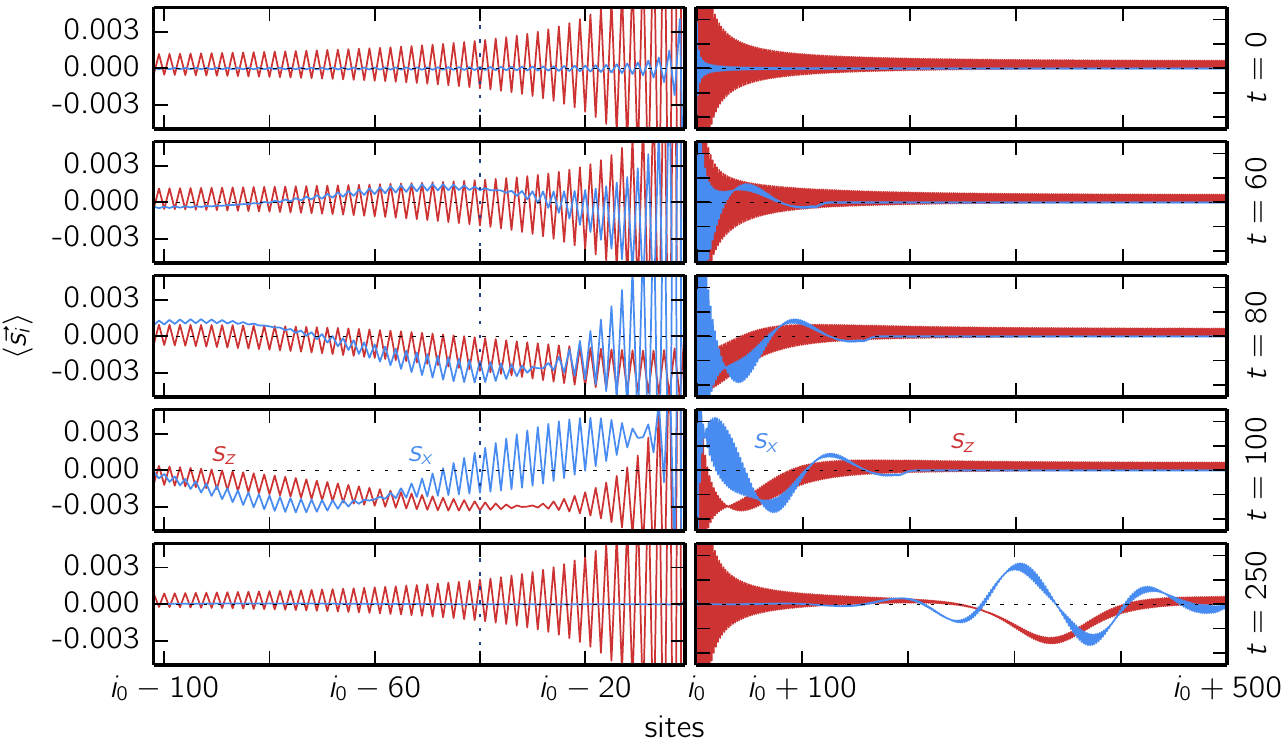}
\caption{(Color online)
Snapshots the of conduction-electron magnetic moments $\langle \ff s_{i} \rangle_{t}$ at different times $t$ as indicated on the right and by the corresponding arrows in Fig.\ \ref{fig:spinden}.
Red lines: $z$-components of $\langle \ff s_{i} \rangle_{t}$.
Blue lines: $x$ components.
The profiles are perfectly symmetric to the impurity site $i=i_{0}$ but displayed up to distances $|i-i_{0}| \le 100$ on the left-hand side and up to the system boundary, $|i-i_{0}| \le 500$, on the right-hand side.
Parameters $J=5, B=0.1$.
} 
\label{fig:cut}
\end{figure*}
%--------------------------------------------------------------------------------------------------------------

Snapshots of the conduction-electron spin dynamics are shown in Fig.\ \ref{fig:cut} for the initial state at $t=0$ and for states at four later times $t>0$ which are also indicated by the arrows in Fig.\ \ref{fig:spinden}.
At $t=0$ the conduction-electron system is in its ground state for the given initial direction of the classical spin. 
The latter basically points into the $-z$ direction, apart from a small positive $x$-component ($\vartheta = \pi /50$) which is  necessary to break the symmetry of the problem and to initiate the dynamics.
This tiny effect will be disregarded in the following.

From the perspective of the conduction-electron system, the interaction term $J \ff S \ff s_{i_0}$ acts as a local external magnetic field $J \ff S$ which locally polarizes the conduction electrons at $i_{0}$.
Since $J$ is antiferromagnetic, the local moment $\langle \ff s_{i_{0}} \rangle$ points into the $+z$ direction. 
At half-filling, the conduction-electron system exhibits pronounced antiferromagnetic spin-spin correlations which give rise to an antiferromagnetic spin-density wave structure aligned to the $z$ axis at $t=0$, see first panel of Fig.\ \ref{fig:cut}.

The total spin $\ff S_{\rm tot} = 0$ at $t=0$, i.e., the classical spin $\ff S$ is exactly compensated by the total conduction-electron spin $\langle \ff s_{\rm tot} \rangle = \sum_{i} \langle \ff s_{i} \rangle = - \ff S$ in the ground state.
This can be traced back to the fact that for a $D=1$-dimensional tight-binding system with an odd number of sites $L$, with $N=L$ and with a single static magnetic impurity, there is exactly one localized state per spin projection $\sigma$, irrespective of the strength of the impurity potential (here given by $J\ff S = 0.5 J \ff e_{z}$). 
The number of $\uparrow$ one-particle eigenstates therefore exceeds the number of $\downarrow$ states by exactly one.

Since the energy of the excitation induced by the external field $\ff B$ is completely dissipated into the bulk, the state of the conduction-electron system at large $t$ (but shorter than $t\approx 500$ where finite-size effects appear) must locally, close to $i_{0}$, resemble the conduction-electron ground state for the reversed spin $\ff S = +0.5 \ff e_{z}$.
This implies that locally all magnetic moments $\langle \ff s_{i} \rangle_{t}$ must reverse their direction.
In fact, the last panel in Fig.\ \ref{fig:cut} (left) shows that the new spin configuration is reached for $t=250$ at sites with distance $|i-i_{0}| \lesssim 100$, see dashed line, for example. 
For later times the spin configuration stays constant (until the wave packet reflected from the system boundaries reaches the vicinity of $i_{0}$).
The reversal is almost perfect, e.g., $\langle \ff s_{i_{0}} \rangle_{t=0} = 0.2649 \to \langle \ff s_{i_{0}} \rangle_{t \ge 250} = -0.2645$.
Deviations of the same order of magnitude are also found at larger distances, e.g., $i=i_{0}-100$.
We attribute those tiny effects to a weak dependence of the local ground state on the non-equilibrium state far from the impurity at $t=250$, see right part of the last panel in Fig.\ \ref{fig:cut}.

The other panels in Fig.\ \ref{fig:cut} demonstrate the mechanism of the spin reversal. 
At short times (see $t=60$, second panel) the perturbation of the initial equilibrium configuration of the conduction-electron moments is still weak.
For $t=80$ and $t=100$ one clearly notices the emission of the wave packet starting. 
Locally, the antiferromagnetic structure is preserved (see left part) but superimposed on this, there is an additional spatial structure of much longer size developing. 
This finally forms the wave packet which is emitted from the central region. 
Its spatial extension is about $\Delta \approx 300$ as can be estimated for $t=250$ (last panel on the right) where it covers the region $200 \lesssim i \lesssim 500$.
The same can be read off from the upper part of Fig.\ \ref{fig:spinden}.
Assuming that the reversal of each of the conduction-electron moments takes about the same time as the reversal of the classical spin, $\Delta$ is roughly given by the reversal time times the Fermi velocity and therefore strongly depends on $J$ and $B$. 
For the present case, we have $\tau_{1} \approx 150 / T$ which implies $\Delta \approx 150 \times 2 = 300$ in rough agreement with the data.

In the course of time, the long-wave length structure superimposed on the short-range antiferromagnetic texture develops a node. 
This can be seen for $t=100$ and $i \approx 40$ (fourth panel, see dashed line). 
The node marks the spatial border between the new (right of the node, closer to $i_{0}$) and the original antiferromagnetic structure of the moments and moves away from $i_{0}$ with increasing time.

At a fixed position $i$, the reversal of the conduction-electron moment $\langle \ff s_{i} \rangle_{t}$ takes place in a similar way as the reversal of the classical spin (see both panels in Fig.\ \ref{fig:spinden} for a fixed $i$).
During the reversal time, its $x$ and $y$ components undergo a precessional motion while the $z$ component changes sign. 
Note, however, that {\em during} the reversal $|\langle \ff s_{i} \rangle|$ gets much larger than its value in the initial and in the final equilibrium state. 

%--------------------------------------------------------------------------------------------------------------
\section{Effective classical spin dynamics}
\label{sec:spinonly}
%--------------------------------------------------------------------------------------------------------------

%--------------------------------------------------------------------------------------------------------------
\subsection{Perturbation theory}
\label{sec:eff}
%--------------------------------------------------------------------------------------------------------------

Eqs.\ (\ref{eq:largecl}) and (\ref{eq:smallcl}) do not form a closed set of equations of motion but must be supplemented by the full equation of motion (\ref{eq:rhoeom}) for the one-particle conduction-electron density matrix.
This implies that the fast electron dynamics must be taken into account explicitly even if the spin dynamics is much slower. 
Hence, there is a strong motivation to integrate out the conduction-electron degrees of freedom altogether and to take advantage from a much larger time step within a corresponding spin-only time-propagation method.
Unfortunately, a simple effective spin-only action can be obtained in the weak-coupling (small-$J$) limit only. \cite{ON06,BNF12}
This weak-coupling approximation is also implicit to all effective spin-only approaches that consider the effect of conduction electrons on the spin dynamics. \cite{ZL04}

In the weak-$J$ limit the electron degrees of freedom can be eliminated in a straightforward way by using standard linear-response theory: \cite{FW71} 
We assume that the initial state at $t=0$ is given by the conduction-electron system in its ground state or in thermal equilibrium and an arbitrary state of the classical spin. 
This may be realized formally by suddenly switching on the interaction $J(t)$ at time $t=0$, i.e., $J(t)=J\Theta(t)$ 
and by switching the local field from some initial value $\ff B_{\rm ini}$ at $t=0$ to a final value $\ff B$ for $t>0$.
The response of the conduction-electron spin at $i_{0}$ and time $t>0$ ($\langle \ff s_{i_{0}} \rangle_{t} = 0$ for $t=0$) due to the time-dependent perturbation $J(t) \ff S(t)$ is
\begin{equation}
  \langle \ff s_{i_{0}} \rangle_{t} =  J \int_{0}^{t} dt' \, \underline{\Pi}^{\rm (ret)}(t,t') \cdot \ff S(t') 
\end{equation}
up to linear order in $J$.
Here, the free ($J=0$) local retarded spin susceptibility of the conduction electrons $\underline{\Pi}^{\rm (ret)}(t,t')$ is a tensor with elements
\begin{equation}
\Pi^{\rm (ret)}_{\alpha\beta}(t,t') = - i \Theta(t-t') \langle [ s^{\alpha}_{i_{0}}(t) , s^{\beta}_{i_{0}}(t') ] \rangle \: ,
\end{equation}
where $\alpha,\beta= x,y,z$.
Using this in Eq.\ (\ref{eq:largecl}), we get an equation of motion for the classical spins only,
\begin{eqnarray}
&& \frac{d}{dt} \ff S(t)  
  = 
\nonumber
 \ff S(t) \times \ff B \\
&& -
 J^{2} \ff S(t) \times    
   \int_{0}^{t} dt' \, \underline{\Pi}^{\rm (ret)}(t-t') \cdot \ff S(t')
\label{eq:retard}   
\end{eqnarray}
which is correct up to order $J^{2}$.
 
This represents an equation of motion for the classical spin only.
It has a temporally non-local structure and includes an effective interaction of the classical spin at time $\ff S(t)$ with the same classical spin at earlier times $t'<t$. 
In the full quantum-classical theory where the electronic degrees of freedom are taken into account exactly, this retarded interaction is mediated by a non-equilibrium electron dynamics starting at site $i_{0}$ and time $t'$ and returning back to the same site $i_{0}$ at time $t>t'$.
Here, for weak $J$, this is replaced by the equilibrium and homogeneous-in-time conduction-electron spin susceptibility $\underline{\Pi}^{\rm (ret)}(t-t')$.
Compared with the results of the full quantum-classical theory, we expect that the perturbative spin-only theory breaks down after a propagation time $t \sim 1/J$ at the latest.

Using Wick's theorem, \cite{FW71} the spin susceptibility is easily expressed in terms of the greater and the lesser equilibrium one-particle Green's functions, $G^{>}_{ii,\sigma\sigma'}(t,t') = - i \langle c_{i\sigma} (t) c^{\dagger}_{i\sigma'} (t') \rangle$ and $G^{<}_{ii,\sigma\sigma'}(t,t') = i \langle c^{\dagger}_{i\sigma'} (t') c_{i\sigma} (t) \rangle$, respectively: 
\begin{eqnarray}
& &
\Pi^{\rm (ret)}_{\alpha\alpha'}(t-t') 
=
\Theta (t-t') \frac{1}{2} 
\nonumber \\
& & 
\times 
\mbox{Im} \:
\tr_{2\times 2} 
\Big[
\sigma^{\alpha} G^{>}_{i_{0}i_{0}}(t,t')
\sigma^{\alpha'} G^{<}_{i_{0}i_{0}}(t',t)
\Big] \; .
\label{eq:pi}
\end{eqnarray}
Assuming that the conduction-electron system is characterized by a real, symmetric and spin-independent hopping matrix $T_{ij}$ (as given by the first term of Eq.\ (\ref{eq:ham})), $G^{>}$ and $G^{<}$ are unit matrices with respect to the spin indices.
They are easily expressed as explicit functions of $\ff T$ (see Ref.\ \onlinecite{BP11}, for example).
With $\tr (\sigma^{\alpha} \sigma^{\alpha'}) = 2\delta_{\alpha\alpha'}$, we find
\begin{eqnarray}
&& \Pi^{\rm (ret)}_{\alpha\alpha'}(t-t')
=
\Theta (t-t')\delta_{\alpha\alpha'} \mbox{Im} 
\nonumber \\
&&\times
\left(
\frac{e^{-i\ff T(t-t')}}{1+e^{-\beta (\ff T - \mu)}}
\right)_{i_{0}i_{0}}
\! \! \!
\left(
\frac{e^{-i\ff T(t'-t)}}{e^{\beta (\ff T - \mu)}+1}
\right)_{i_{0}i_{0}}
\label{eq:pifinal}
\end{eqnarray}
for a conduction-electron system at inverse temperature $\beta$ and chemical potential $\mu$.

%--------------------------------------------------------------------------------------------------------------
\begin{figure}[t]
\centering
\includegraphics[width=\columnwidth]{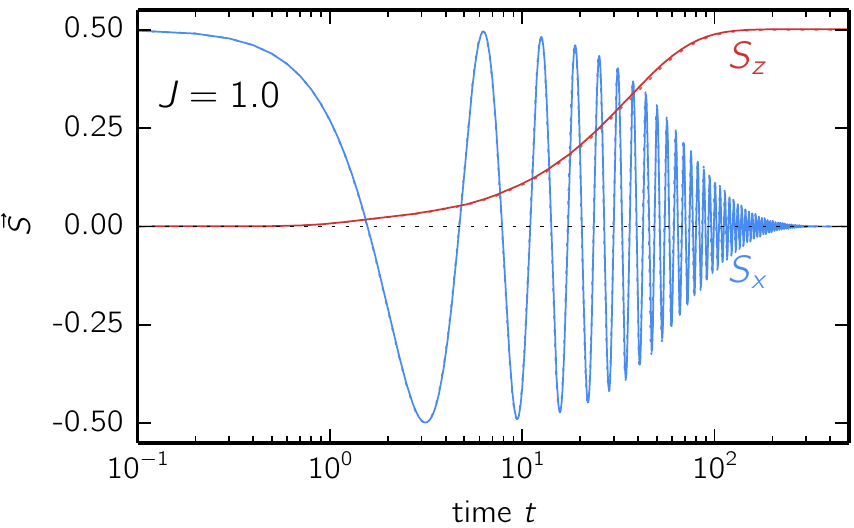}
\caption{(Color online)
Components $S_{x}(t)$ and $S_{z}(t)$ of the classical spin after a sudden switch of the field from $x$ to $z$ direction at $t=0$. 
Calculations for $B=1$ and $J=1$.
Solid lines: results of the linear-response dynamics, Eq.\ (\ref{eq:retard}). 
Dashed lines: results of the exact quantum-classical dynamics for 
$L=1001$. 
} 
\label{fig:lra}
\end{figure}
%-------------------------------------------------------------------------------------------------------------- 

Using Eq.\ (\ref{eq:pifinal}) we have computed the spin susceptibility for the ground state ($\beta=\infty$) of the conduction-electron system.
This fixes the kernel in the integro-differential equation Eq.\ (\ref{eq:retard}) which is solved numerically by standard techniques. \cite{PTVF07}
We again consider the system displayed in Fig.\ \ref{fig:system} with a single classical spin coupled via $J$ to the central site $i_{0}$ of a chain consisting of $L=1001$ sites. 
Finite-size artifacts do not show up before $t_{\rm max} = 500$.

\begin{figure}[t]
\centering
\includegraphics[width=\columnwidth]{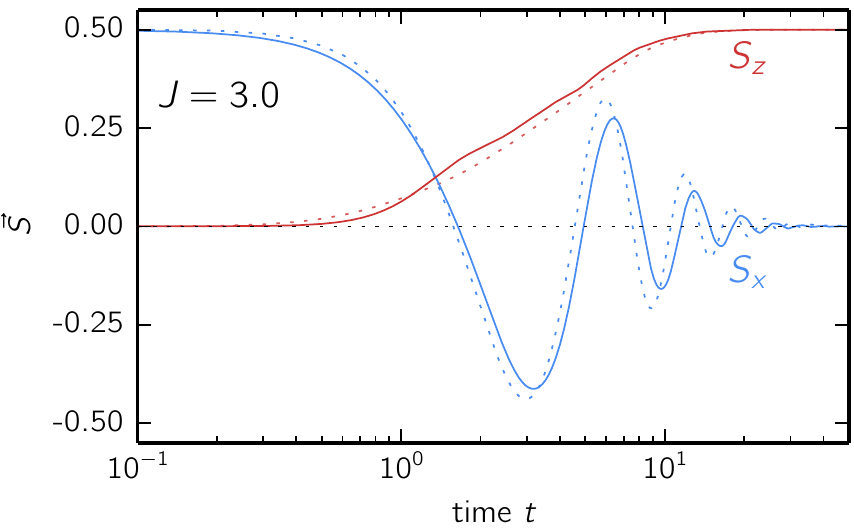}
\caption{(Color online)
The same as Fig.\ \ref{fig:lra} but for a different exchange coupling constant $J=3$. 
} 
\label{fig:lrb}
\end{figure}

Figs.\ \ref{fig:lra} and \ref{fig:lrb} show the resulting linear-response dynamics of the classical spin after preparing the initial state of the system with the classical spin pointing into $+x$ direction while $\ff B = (0,0,B)$.
The external magnetic field induces a precessional motion of the classical spin: there is a rapid oscillation of its $x$ component (and of its $y$ component, not shown) with frequency $\omega \approx B$ (blue lines).
Damping is induced by dissipation of energy and spin: for large times, the $z$ component aligns to the external field (red lines). 

For weak coupling, up to $J=1$ (Fig.\ \ref{fig:lra}), there is an almost perfect agreement between the results of the exact quantum-classical dynamics (full lines) and the linear-response theory (dashed lines) up to the maximum propagation time $t_{\rm max} = 500$. 
We note that, compared to the full theory, there is a tiny deviation of the linear-response result for the $z$ component of $\ff S(t)$ visible in Fig.\ \ref{fig:lra} for times $t \gtrsim 10$.
Hence, on this level of accuracy, $t_1 \approx 10$ sets the time scale up to which the linear-response theory is valid. 
This may appear surprising as this implies $t_{1} \, J = 10$ for the ``small'' dimensionless parameter of the perturbation theory. 
One has to keep in mind, however, that even if the perturbation is ``strong'', its {\em effects} can be rather moderate since 
only non-adiabatic terms $\sim \ff S(t) \times \ff S(t')$ contribute in Eq.\ (\ref{eq:retard}).
 
%--------------------------------------------------------------------------------------------------------------
%--------------------------------------------------------------------------------------------------------------

For $J=3$, see Fig.\ \ref{fig:lrb}, damping of the classical spin sets in much earlier. 
Visible deviations of the linear-response theory from the full dynamics already appear on a time scale that is almost two orders of magnitude smaller as compared to the case $J=1$. 
A simple reasoning based on the argument that the dimensionless expansion parameter is $t_{1} \, J$ fails as this disregards the strong enhancement of retardation effects with increasing $J$, which have been discussed in Sec.\ \ref{sec:rel}.
These effects make the perturbation much more effective, i.e., lead to a torque, which is exerted by the conduction electrons on the classical spin, growing stronger than linear in $J$.

We conclude that linear-response theory is highly attractive formally as it provides a tractable spin-only effective theory. 
On the other hand, substantial discrepancies compared to the full (non-perturbative) theory show up as soon as damping effects become stronger. 
Note that, with increasing time, these deviations must diminish and disappear eventually since both, the full and the effective theory, predict a fully relaxed spin state for $t \to \infty$ -- see lower panel of Fig.\ \ref{fig:lrb}, for example. 
At least for simple systems with a single classical spin, as considered here, this implies that the effective theory provides qualitatively reasonable results.

%--------------------------------------------------------------------------------------------------------------
\subsection{Landau-Lifschitz-Gilbert equation}
\label{sec:llg}
%--------------------------------------------------------------------------------------------------------------

To derive the LLG equation, the linear-response theory must be further simplified: \cite{BNF12,Fra08}
As is obvious from Eq.\ (\ref{eq:retard}), the spin-susceptibility $\underline{\Pi}^{\rm (ret)}(t-t')$ can be interpreted as an effective retarded self-interaction of the spin. 
We assume that the electron dynamics is much faster than the spin dynamics.
On the time scale of the spin dynamics, the self-interaction then takes place almost instantaneously, i.e., the memory kernel $\Pi_{\alpha\alpha'}^{\rm (ret)}(t-t') = \delta_{\alpha\alpha'} \Pi^{\rm (ret)}(t-t')$ in Eq.\ (\ref{eq:retard}) is peaked at $t' \approx t$.
We can therefore approximate $\ff S(t') \approx \ff S(t) + (t'-t) \dot{\ff S}(t)$ under the integral in Eq.\ (\ref{eq:retard}). 
This immediately yields:
\begin{equation}
  \frac {d \ff S(t)} {dt}
  =
  \ff S(t) \times \ff B
  +
  \alpha(t) \, \ff S(t) \times \frac {d \ff S(t)} {dt} \: ,
\label{eq:alphat}
\end{equation}
where
\begin{equation}
  \alpha(t) 
  = 
  J^2 \int_{0}^{t} d\tau \, \tau \, \Pi^{\rm (ret)}(\tau) 
\end{equation}
after substituting $t' \mapsto \tau=t-t'$.

Eq.\ (\ref{eq:alphat}) takes the form of the standard LLG equation for a single classical spin [cf.\ Eq.\ (\ref{eq:llg})] if $\alpha(t)$ is replaced by 
\begin{equation}
  \alpha \equiv \lim_{t\to \infty} \alpha(t) = J^2 \int_{0}^{\infty} dt \, t \, \Pi^{\rm (ret)}(t) \: .
\label{eq:gil}  
\end{equation}
Note that this is a necessary step to arrive at a {\em constant} damping parameter which can again be justified by noting that $\Pi^{\rm (ret)}(t)$ is peaked at $t=0$.

Before proceeding, let us stress that Eq.\ (\ref{eq:gil}) is, or is equivalent to, the standard expression used for computing the Gilbert damping constant in various studies: 
After Fourier transformation, 
\begin{equation}
  \Pi^{\rm (ret)}(\omega) = \int dt\, e^{i\omega t} \,  \Pi^{\rm (ret)}(t) \: , 
\label{eq:fourier}  
\end{equation}
one ends up with 
\begin{equation}
  \alpha 
  =  - i J^2 \frac{\partial}{\partial \omega} \Pi^{\rm (ret)}(\omega) \Big|_{\omega=0} 
  = J^2 \frac{\partial}{\partial \omega} \mbox{Im} \, \Pi^{\rm (ret)}(\omega) \Big|_{\omega=0} 
  \: ,
\label{eq:gil2}  
\end{equation}
which has also been derived, e.g., in Refs.\ \onlinecite{SH03,BNF12} in different contexts. 
The frequency-dependent spin correlation $\Pi^{\rm (ret)}(\omega)$ in Eq.\ (\ref{eq:gil2}) can be obtained explicitly as the Fourier transform of $\Pi^{\rm (ret)}(t)$ given by Eq.\ (\ref{eq:pifinal}).
A straightforward calculation yields:
\begin{equation}
  \alpha 
  = 
  \frac{\pi}{2} J^{2} \int d\omega \frac{d f(\omega)}{d\omega} \, A_{\rm loc}(\omega) A_{\rm loc}(\omega)
  \: ,
\label{eq:gil3}  
\end{equation}
where $f(\omega) = 1/(\exp(\beta \omega) +1)$ is the Fermi function, and 
\begin{equation}
  A_{\rm loc}(\omega) = L^{-1} \sum_{\ff k} \delta(\omega + \mu - \varepsilon(\ff k))
\label{eq:spden}
\end{equation} 
for $L\to \infty$ is the local one-particle spectral function.
Here we have assumed periodic boundary conditions, i.e., spatial homogeneity, such that the hopping matrix $\ff T$ is diagonalized by Fourier transformation:
\begin{equation}
  T_{ij} = \sum_{\ff k} U_{i\ff k} \varepsilon(\ff k) U^{\dagger}_{\ff k j} 
\end{equation}
with $U_{i\ff k} = L^{-1/2} \exp(i \ff k \ff R_{i})$. 
The eigenvalues of $\ff T$ are given by the tight-binding dispersion $\varepsilon(\ff k)$ of the conduction-electron Bloch band.

Within our simple tight-binding model for the conduction-electron system, Eqs.\ (\ref{eq:gil2}) and (\ref{eq:gil3}) are equivalent with Kambersky's breathing Fermi-surface theory, related torque-correlation models and scattering theory and have frequently been used for {\em ab initio} as well as model computations of the Gilbert damping constant. \cite{KK02,Kam07,BTB08,EMKK11,Sak12,UMS12,MKWE13,TH14} 

Let us remark that Eq.\ (\ref{eq:gil3}) demonstrates that $\alpha<0$.
This results from the convention for the coupling of the magnetic field to the spin, namely $H = H(\ff B=0) - \ff B \ff S$ [see Eq.\ \ref{eq:ham})], which has been adopted here. 
As a consequence, the precession of $\ff S(t)$ around $\ff B$ is described by a {\em left}-hand helix, $\dot{\ff S} = \ff S(t) \times \ff B$, and thus $\alpha$ must be negative to describe damping.

%--------------------------------------------------------------------------------------------------------------
\subsection{Ill-defined Gilbert damping}
\label{sec:ill}
%--------------------------------------------------------------------------------------------------------------

The above discussion shows that Eq.\ (\ref{eq:gil}) represents the fundamental definition of the Gilbert damping constant $\alpha$ and that the limit $t \to \infty$ is crucial to recover the LLG equation in its standard form. 
The existence of the long-time limit, however, decisively depends on the long-time behavior of the retarded spin-correlation function $\Pi^{\rm (ret)}(t)$. 
Starting from Eq.\ (\ref{eq:pifinal}), this is easily computed as
\begin{equation}
\Pi^{\rm (ret)}(t)
=
\Theta (t) \mbox{Im} \frac{1}{L^{2}}
\sum_{\ff k,\ff p}
\frac{e^{-i\varepsilon(\ff k)t}}{1+e^{-\beta (\varepsilon(\ff k) - \mu)}}
\frac{e^{i\varepsilon(\ff p)t}}{e^{\beta (\varepsilon(\ff p) - \mu)}+1}
\label{eq:pieps}
\: .
\end{equation}

%--------------------------------------------------------------------------------------------------------------
\begin{figure}[t]
\centering
\includegraphics[width=\columnwidth]{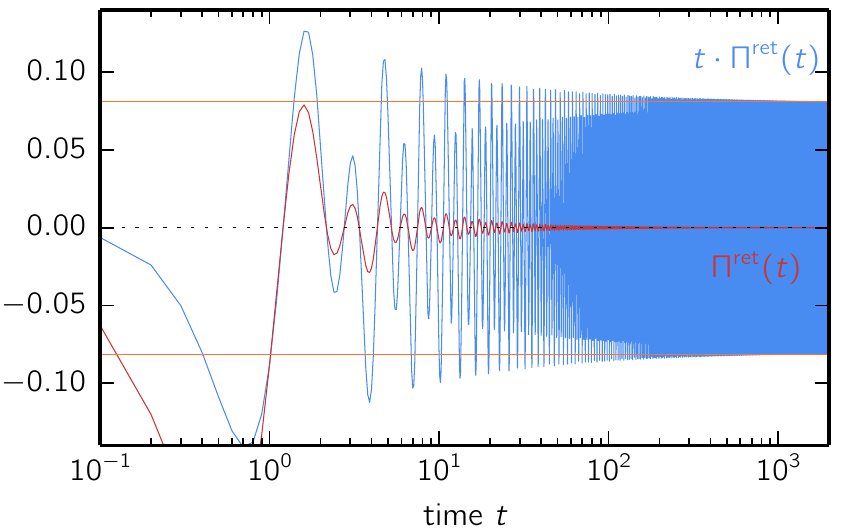}
\caption{(Color online)
Local retarded spin susceptibility of the conduction electrons $\Pi^{\rm (ret)}(t)$ (red line) and $t\,\Pi^{\rm (ret)}(t)$ (blue) as obtained from Eq.\ (\ref{eq:pieps}) for a one-dimensional conduction-electron system with $L=10000$ sites and periodic boundary conditions.
} 
\label{fig:pilong}
\end{figure}
%--------------------------------------------------------------------------------------------------------------

Fig.\ \ref{fig:pilong} gives an example for the time-dependence of $\Pi^{\rm (ret)}(t)$. 
The calculations have been done at half-filling, $\beta=\infty$ for $L=10^{4}$ sites.
We note that the susceptibility is in fact peaked at $t = 0$. 
For long times, it oscillates with frequency $\omega_{\Pi} = 4$ and decays as $1/t$. 
This is an important observation as it implies that the limit in Eq.\ (\ref{eq:gil}) does not exist and that, therefore, the damping constant $\alpha$ is ill-defined.

To analyze the physical origin of the divergent integral, we rewrite the spin susceptibility in Eq.\ (\ref{eq:pieps}) as
\begin{equation}
\Pi^{\rm (ret)}(t)
=
\Theta (t) \mbox{Im} 
[
A_{\rm loc}^{\rm (unocc)}(-t) 
A_{\rm loc}^{\rm (occ)}(t) 
] \: .
\label{eq:pi1}
\end{equation}
Its long-time behavior is governed by  the long-time behavior of the Fourier transform
\begin{equation}
A_{\rm loc}^{\rm (occ, unocc)} (t) = \int d\omega \, e^{i\omega t} A_{\rm loc}^{\rm (occ, unocc)} (\omega)
\end{equation}
of the occupied,
$A_{\rm loc}^{\rm (occ)}(\omega) \equiv f(\omega) A_{\rm loc}(\omega)$, 
and of the unoccupied,
$A_{\rm loc}^{\rm (unocc)}(\omega) \equiv f(\omega) A_{\rm loc}(\omega)$,
part of the spectral density, see Eq.\ (\ref{eq:spden}).

For functions with smooth $\omega$ dependence, the Fourier transform generically drops to zero exponentially fast if $t\to \infty$. 
A power-law decay, however, is obtained if there are singularities of $A_{\rm loc}^{\rm (occ, unocc)}(\omega)$.
We can distinguish between van Hove singularities, which are, e.g., of the form $\propto \Theta(\omega-\omega_{0})(\omega - \omega_{0})^{k}$ (with $k > -1$), and the step-like singularity $\propto \Theta(\omega - \omega_{0})$ (i.e.\ $k=0$), arising in the zero-temperature limit at $\omega_{0} = 0$ due to the Fermi function.
Generally, a singularity of order $k$ gives rise to the asymptotic behavior $A_{\rm loc}^{\rm (occ, unocc)} (t) \propto t^{-1-k}$, apart from a purely oscillatory factor $e^{i\omega_{0}t}$. 
For the present case, the van Hove singularities of $A_{\rm loc}^{\rm (occ, unocc)}(\omega)$ at $\pm \omega_{0} = 2$ explain, via Eq.\ (\ref{eq:pi1}) the oscillation of $\Pi^{\rm (ret)}(t)$ with frequency $\omega_{\Pi}=2\omega_{0}=4$.

Generally, the location of the van Hove singularity on the frequency axis, i.e.\ $\omega_{0}$, determines the oscillation period while the decay of $\Pi^{\rm (ret)}(t)$ is governed by the strength of the singularity. 
Consider, as an example, the zero-temperature case and assume that there are no van Hove singularities. 
The sharp Fermi edge implies $A_{\rm loc}^{\rm (occ, unocc)} (t) \propto t^{-1}$, and thus $\Pi^{\rm (ret)}(t) \propto t^{-2}$.
The Gilbert-damping constant is well defined in this case.

The strength of van Hove singularities depends on the lattice dimension $D$. \cite{AM76}
For a one-dimensional lattice, we have van Hove singularities with $k=-1/2$, and thus $\Pi^{\rm (ret)}(t) \propto t^{-1}$, consistent with Fig.\ (\ref{fig:pilong}).
Here, the strong van Hove singularity dominates the long-time asymptotic behavior as compared to the weaker Fermi-edge singularity.
For $D=3$, we have $k=1/2$ and $\Pi^{\rm (ret)}(t) \propto t^{-3}$ if $\beta<\infty$ while for $\beta=\infty$ the Fermi-edge dominates and $\Pi^{\rm (ret)}(t) \propto t^{-2}$.
The $D=2$ case is more complicated:
The logarithmic van Hove singularity $\propto \ln |\omega|$ leads to $\Pi^{\rm (ret)}(t) \propto t^{-2}$.
This, however, applies to cases off half-filling only. 
At half-filling the van Hove and the Fermi-edge singularity combine to a singularity $\propto \Theta(\omega) \ln |\omega|$ which gives $\Pi^{\rm (ret)}(t) \propto \ln^{2}(t) / t^{2}$. 
For finite temperatures, we again have $\Pi^{\rm (ret)}(t) \propto t^{-2}$.

The existence of the integral Eq.\ (\ref{eq:gil}) depends on the $t\to \infty$ behavior and either requires a decay as $\Pi^{\rm (ret)}(t) \propto t^{-3}$ or faster, or an asymptotic form $\Pi^{\rm (ret)}(t) \propto e^{i\omega_{0}t} / t^{2}$ with an oscillating factor resulting from a non-zero position $\omega_{0} \ne 0$ of the van Hove singularity.
For the one-dimensional case, we conclude that the LLG equation (with a time-independent damping constant) is based on an ill-defined concept.
Also, the derivation of Eqs.\ (\ref{eq:gil2}) and (\ref{eq:gil3}) is invalid in this case as the $\omega$ derivative and the $t$ integral do not commute.
This conclusion might change for the case of {\em interacting} conduction electrons.
Here one would expect a regularization of van Hove singularities due to a finite imaginary part of the conduction-electron self-energy. 

%--------------------------------------------------------------------------------------------------------------
\section{Conclusions}
\label{sec:con}
%--------------------------------------------------------------------------------------------------------------

Hybrid systems consisting of classical spins coupled to a bath of non-interacting conduction electrons represent a class of model systems with a non-trivial real-time dynamics which is numerically accessible on long time scales.
Here we have considered the simplest variant of this class, the Kondo-impurity model with a classical spin, and studied the relaxation dynamics of the spin in an external magnetic field.
As a fundamental model this is interesting of its own but also makes contact with different fields, e.g., atomistic spin dynamics in magnetic samples, 
spin relaxation in spintronics devices, 
femto-second dynamics of highly excited electron systems where local magnetic moments are formed due to electron correlations,
and artificial Kondo systems simulated with ultracold atoms in optical lattices.

We have compared the coupled spin and electron dynamics with the predictions of the widely used Landau-Lifshitz-Gilbert equation which is supposed to cover the regime of weak local exchange $J$ and slow spin dynamics.
For the studied setup, the LLG equation predicts a rather regular time evolution characterized by spin precession, spin relaxation and eventually reversal of the spin on a time scale $\tau$ depending on $J$ (and the field strength $B$).
We have demonstrated that this type of dynamics can be recovered and understood on a microscopic level in the more fundamental quantum-classical Kondo model.
It is traced back to a non-adiabatic dynamics of the electron degrees of freedom and the feedback of the electronic subsystem on the spin.
It turns out that the spin dynamics is essentially a consequence of the retarded effect of the local exchange. 
Namely, the classical spin can be seen as a perturbation exciting the conduction-electron system locally.
This electronic excitation propagates and feeds back to the classical spin, but at a later time, and thereby induces a spin torque.

We found that this mechanism drives the relaxation of the system to its local ground state irrespective of the strength of the local exchange $J$.
As the microscopic dynamics is fully conserving, the energy and spin of the initial excitation  which is locally stored in the vicinity of the classical spin, must be dissipated into the bulk of the system in the course of time.
This dissipation could be uncovered by studying the relaxation process from the perspective of the electron degrees of freedom.
Dissipation of energy and spin takes place through the emission of a dispersive spin-polarized wave packet propagating through the lattice with the Fermi velocity.
In this process the local conduction-electron magnetic moment at any given distance to the impurity undergoes a reversal, characterized by precession and relaxation, similar to the motion of the classical spin.

The dynamics of the classical spin can be qualitatively very different from the predictions of the LLG equation for strong $J$.
In this regime we found a complex motion characterized by oscillations of the angle between the classical spin $\ff S(t)$ and the local conduction-electron magnetic moment at the impurity site $\langle \ff s_{i_{0}} \rangle$ around the adiabatic value $\gamma = \pi$ which takes place on an emergent new time scale.

In the weak-$J$ limit, the classical spin dynamics is qualitatively predicted correctly by the LLG equation.
At least partially, however, this must be attributed to the fact that the LLG approach, by construction, recovers the correct {\em final} state where the spin is parallel to the field.
In fact, quantitative deviations are found {\em during} the relaxation process.
The LLG approach is based on first-order perturbation theory in $J$ and on the additional assumption that the classical spin is slow. 
To pinpoint the source of the deviations, we have numerically solved the integro-differential equation that is obtained in first-order-in-$J$ perturbation theory and compared with the full hybrid dynamics.
The deviations of the perturbative approach from the exact dynamics are found to gradually  increase with the propagation time (until the proximity to the final state enforces the correct long-time asymptotics).
This is the expected result as the dimensionless small parameter is $J t$.
However, with increasing $J$ the time scale on which perturbation theory is reliable decreases much stronger than $1/J$ due to a strong enhancement of retardation effects which make the perturbation more effective and produce a stronger torque.

Generally, the perturbation can be rather {\em ineffective} in the sense that it produces a torque $\propto \ff S(t) \times \ff S(t')$ which is very weak if the process is nearly adiabatic.
This explains that first-order perturbation theory and the LLG equation is applicable at all for couplings of the order of hopping $J \sim T$.
For the present study this can also be seen as a fortunate circumstance since the regime of very weak couplings $J \ll T$ is not accessible numerically.
In this case the spin-reversal time scale gets so large that the propagation of excitations in the conduction-electron subsystem would by affected by backscattering from the edges of the system which necessarily must be assumed as finite for the numerical treatment.

For the one-dimensional lattice studied here, a direct comparison between LLG equation and the exact quantum-classical theory is not meaningful as the damping constant $\alpha$ is ill-defined in this case.
We could argue that the problem results from the strength of the van Hove singularities in the conduction-electron density of states which dictates the long-time behavior of the memory kernel of the integro-differential equation which is given by the equilibrium spin susceptibility.
As the type of the van Hove singularity is characteristic for all systems of a given dimension, we can generally conclude that the LLG approach reduces to a purely phenomenological scheme in the one-dimensional case.
However, it is an open question, which will be interesting to tackle in the future, if this conclusion is still valid for systems where the Coulomb interaction among the conduction electrons is taken into account additionally.

There are more interesting lines of research which are based on the present work and could be pursued in the future.
Those include systems with more than a single spin where, e.g., the effects of a time-dependent and retarded RKKY interaction can be studied additionally.
We are also working on a tractable extension of the theory to account for longitudinal fluctuations of the spins to include time-dependent Kondo screening, and the competition with RKKY coupling, on a time-dependent mean-field level.
Finally, lattice rather than impurity variants of the quantum-classical hybrid model are highly interesting to address the time-dependent phase transitions.
\\

%--------------------------------------------------------------------------------------------------------------
\acknowledgments
%--------------------------------------------------------------------------------------------------------------

We would like to thank M.\ Eckstein, A.\ Lichtenstein, R.\ Rausch, E.\ Vedmedenko and R.\ Walz for instructive discussions.
Support of this work by the Deutsche Forschungsgemeinschaft within the SFB 668 (project B3) and within the SFB 925 (project B5) is gratefully acknowledged.


\begin{thebibliography}{53}
\expandafter\ifx\csname natexlab\endcsname\relax\def\natexlab#1{#1}\fi
\expandafter\ifx\csname bibnamefont\endcsname\relax
  \def\bibnamefont#1{#1}\fi
\expandafter\ifx\csname bibfnamefont\endcsname\relax
  \def\bibfnamefont#1{#1}\fi
\expandafter\ifx\csname citenamefont\endcsname\relax
  \def\citenamefont#1{#1}\fi
\expandafter\ifx\csname url\endcsname\relax
  \def\url#1{\texttt{#1}}\fi
\expandafter\ifx\csname urlprefix\endcsname\relax\def\urlprefix{URL }\fi
\providecommand{\bibinfo}[2]{#2}
\providecommand{\eprint}[2][]{\url{#2}}

\bibitem[{\citenamefont{Landau and Lifshitz}(1935)}]{LL}
\bibinfo{author}{\bibfnamefont{L.~D.} \bibnamefont{Landau}} \bibnamefont{and}
  \bibinfo{author}{\bibfnamefont{E.~M.} \bibnamefont{Lifshitz}},
  \bibinfo{journal}{Phys. Z. Sow.} \textbf{\bibinfo{volume}{153}}
  (\bibinfo{year}{1935}).

\bibitem[{\citenamefont{Gilbert}(1955)}]{LLG}
\bibinfo{author}{\bibfnamefont{T.}~\bibnamefont{Gilbert}},
  \bibinfo{journal}{Phys. Rev.} \textbf{\bibinfo{volume}{100}},
  \bibinfo{pages}{1243} (\bibinfo{year}{1955}).

\bibitem[{\citenamefont{Gilbert}(2004)}]{1353448}
\bibinfo{author}{\bibfnamefont{T.}~\bibnamefont{Gilbert}},
  \bibinfo{journal}{Magnetics, IEEE Transactions on}
  \textbf{\bibinfo{volume}{40}}, \bibinfo{pages}{3443} (\bibinfo{year}{2004}).

\bibitem[{\citenamefont{Aharoni}(1996)}]{Aharoni}
\bibinfo{author}{\bibfnamefont{A.}~\bibnamefont{Aharoni}},
  \emph{\bibinfo{title}{Introduction to the Theory of Ferromagnetism}}
  (\bibinfo{publisher}{Oxford University Press}, \bibinfo{address}{Oxford},
  \bibinfo{year}{1996}).

\bibitem[{\citenamefont{Tatara et~al.}(2008)\citenamefont{Tatara, Kohno, and
  Shibata}}]{TKS08}
\bibinfo{author}{\bibfnamefont{G.}~\bibnamefont{Tatara}},
  \bibinfo{author}{\bibfnamefont{H.}~\bibnamefont{Kohno}}, \bibnamefont{and}
  \bibinfo{author}{\bibfnamefont{J.}~\bibnamefont{Shibata}},
  \bibinfo{journal}{Physics Reports} \textbf{\bibinfo{volume}{468}},
  \bibinfo{pages}{213} (\bibinfo{year}{2008}).

\bibitem[{\citenamefont{Skubic et~al.}(2008)\citenamefont{Skubic, Hellsvik,
  Nordstr\"om, and Eriksson}}]{SHNE08}
\bibinfo{author}{\bibfnamefont{B.}~\bibnamefont{Skubic}},
  \bibinfo{author}{\bibfnamefont{J.}~\bibnamefont{Hellsvik}},
  \bibinfo{author}{\bibfnamefont{L.}~\bibnamefont{Nordstr\"om}},
  \bibnamefont{and} \bibinfo{author}{\bibfnamefont{O.}~\bibnamefont{Eriksson}},
  \bibinfo{journal}{J. Phys.: Condens. Matter} \textbf{\bibinfo{volume}{20}},
  \bibinfo{pages}{315203} (\bibinfo{year}{2008}).

\bibitem[{\citenamefont{Bertotti et~al.}(2009)\citenamefont{Bertotti,
  Mayergoyz, and Serpico}}]{BMS09}
\bibinfo{author}{\bibfnamefont{G.}~\bibnamefont{Bertotti}},
  \bibinfo{author}{\bibfnamefont{I.~D.} \bibnamefont{Mayergoyz}},
  \bibnamefont{and} \bibinfo{author}{\bibfnamefont{C.}~\bibnamefont{Serpico}},
  \emph{\bibinfo{title}{Nonlinear Magnetization Dynamics in Nanosystemes}}
  (\bibinfo{publisher}{Elsevier}, \bibinfo{address}{Amsterdam},
  \bibinfo{year}{2009}).

\bibitem[{\citenamefont{F\"ahnle and Illg}(2011)}]{FI11}
\bibinfo{author}{\bibfnamefont{M.}~\bibnamefont{F\"ahnle}} \bibnamefont{and}
  \bibinfo{author}{\bibfnamefont{C.}~\bibnamefont{Illg}}, \bibinfo{journal}{J.
  Phys.: Condens. Matter} \textbf{\bibinfo{volume}{23}},
  \bibinfo{pages}{493201} (\bibinfo{year}{2011}).
  
\bibitem[{\citenamefont{Evans et~al.}(2014)\citenamefont{Evans, Fan,
  Chureemart, Ostler, Ellis, and Chantrell}}]{EFC+14}
\bibinfo{author}{\bibfnamefont{R.~F.~L.} \bibnamefont{Evans}},
  \bibinfo{author}{\bibfnamefont{W.~J.} \bibnamefont{Fan}},
  \bibinfo{author}{\bibfnamefont{P.}~\bibnamefont{Chureemart}},
  \bibinfo{author}{\bibfnamefont{T.~A.} \bibnamefont{Ostler}},
  \bibinfo{author}{\bibfnamefont{M.~O.~A.} \bibnamefont{Ellis}},
  \bibnamefont{and} \bibinfo{author}{\bibfnamefont{R.~W.}
  \bibnamefont{Chantrell}}, \bibinfo{journal}{J. Phys.: Condens. Matter}
  \textbf{\bibinfo{volume}{26}}, \bibinfo{pages}{103202}
  (\bibinfo{year}{2014}).

\bibitem[{\citenamefont{Ruderman and Kittel}(1954)}]{RK54}
\bibinfo{author}{\bibfnamefont{M.~A.} \bibnamefont{Ruderman}} \bibnamefont{and}
  \bibinfo{author}{\bibfnamefont{C.}~\bibnamefont{Kittel}},
  \bibinfo{journal}{Phys. Rev.} \textbf{\bibinfo{volume}{96}},
  \bibinfo{pages}{99} (\bibinfo{year}{1954}).

\bibitem[{\citenamefont{Kasuya}(1956)}]{Kas56}
\bibinfo{author}{\bibfnamefont{T.}~\bibnamefont{Kasuya}},
  \bibinfo{journal}{Prog. Theor. Phys.} \textbf{\bibinfo{volume}{16}},
  \bibinfo{pages}{45} (\bibinfo{year}{1956}).

\bibitem[{\citenamefont{Yosida}(1957)}]{Yos57}
\bibinfo{author}{\bibfnamefont{K.}~\bibnamefont{Yosida}},
  \bibinfo{journal}{Phys. Rev.} \textbf{\bibinfo{volume}{106}},
  \bibinfo{pages}{893} (\bibinfo{year}{1957}).

\bibitem[{\citenamefont{Onoda and Nagaosa}(2006)}]{ON06}
\bibinfo{author}{\bibfnamefont{M.}~\bibnamefont{Onoda}} \bibnamefont{and}
  \bibinfo{author}{\bibfnamefont{N.}~\bibnamefont{Nagaosa}},
  \bibinfo{journal}{Phys. Rev. Lett.} \textbf{\bibinfo{volume}{96}},
  \bibinfo{pages}{066603} (\bibinfo{year}{2006}).

\bibitem[{\citenamefont{Bhattacharjee et~al.}(2012)\citenamefont{Bhattacharjee,
  Nordstr\"om, and Fransson}}]{BNF12}
\bibinfo{author}{\bibfnamefont{S.}~\bibnamefont{Bhattacharjee}},
  \bibinfo{author}{\bibfnamefont{L.}~\bibnamefont{Nordstr\"om}},
  \bibnamefont{and} \bibinfo{author}{\bibfnamefont{J.}~\bibnamefont{Fransson}},
  \bibinfo{journal}{Phys. Rev. Lett.} \textbf{\bibinfo{volume}{108}},
  \bibinfo{pages}{057204} (\bibinfo{year}{2012}).

\bibitem[{\citenamefont{Umetsu et~al.}(2012)\citenamefont{Umetsu, Miura, and
  Sakuma}}]{UMS12}
\bibinfo{author}{\bibfnamefont{N.}~\bibnamefont{Umetsu}},
  \bibinfo{author}{\bibfnamefont{D.}~\bibnamefont{Miura}}, \bibnamefont{and}
  \bibinfo{author}{\bibfnamefont{A.}~\bibnamefont{Sakuma}},
  \bibinfo{journal}{J. Appl. Phys.} \textbf{\bibinfo{volume}{111}},
  \bibinfo{eid}{07D117} (\bibinfo{year}{2012}).

\bibitem[{\citenamefont{Antropov et~al.}(1995)\citenamefont{Antropov,
  Katsnelson, van Schilfgaarde, and Harmon}}]{AKvSH95}
\bibinfo{author}{\bibfnamefont{V.~P.} \bibnamefont{Antropov}},
  \bibinfo{author}{\bibfnamefont{M.~I.} \bibnamefont{Katsnelson}},
  \bibinfo{author}{\bibfnamefont{M.}~\bibnamefont{van Schilfgaarde}},
  \bibnamefont{and} \bibinfo{author}{\bibfnamefont{B.~N.}
  \bibnamefont{Harmon}}, \bibinfo{journal}{Phys. Rev. Lett.}
  \textbf{\bibinfo{volume}{75}}, \bibinfo{pages}{729} (\bibinfo{year}{1995}).

\bibitem[{\citenamefont{Antropov et~al.}(1996)\citenamefont{Antropov,
  Katsnelson, Harmon, van Schilfgaarde, and Kusnezov}}]{AKH+96}
\bibinfo{author}{\bibfnamefont{V.~P.} \bibnamefont{Antropov}},
  \bibinfo{author}{\bibfnamefont{M.~I.} \bibnamefont{Katsnelson}},
  \bibinfo{author}{\bibfnamefont{B.~N.} \bibnamefont{Harmon}},
  \bibinfo{author}{\bibfnamefont{M.}~\bibnamefont{van Schilfgaarde}},
  \bibnamefont{and} \bibinfo{author}{\bibfnamefont{D.}~\bibnamefont{Kusnezov}},
  \bibinfo{journal}{Phys. Rev. B} \textbf{\bibinfo{volume}{54}},
  \bibinfo{pages}{1019} (\bibinfo{year}{1996}).

\bibitem[{\citenamefont{Kune\v{s} and Kambersk\'y}(2002)}]{KK02}
\bibinfo{author}{\bibfnamefont{J.}~\bibnamefont{Kune\v{s}}} \bibnamefont{and}
  \bibinfo{author}{\bibfnamefont{V.}~\bibnamefont{Kambersk\'y}},
  \bibinfo{journal}{Phys. Rev. B} \textbf{\bibinfo{volume}{65}},
  \bibinfo{pages}{212411} (\bibinfo{year}{2002}).

\bibitem[{\citenamefont{Capelle and Gyorffy}(2003)}]{CG03}
\bibinfo{author}{\bibfnamefont{K.}~\bibnamefont{Capelle}} \bibnamefont{and}
  \bibinfo{author}{\bibfnamefont{B.~L.} \bibnamefont{Gyorffy}},
  \bibinfo{journal}{Europhys. Lett.} \textbf{\bibinfo{volume}{61}},
  \bibinfo{pages}{354} (\bibinfo{year}{2003}).

\bibitem[{\citenamefont{Ebert et~al.}(2011)\citenamefont{Ebert, Mankovsky,
  K\"odderitzsch, and Kelly}}]{EMKK11}
\bibinfo{author}{\bibfnamefont{H.}~\bibnamefont{Ebert}},
  \bibinfo{author}{\bibfnamefont{S.}~\bibnamefont{Mankovsky}},
  \bibinfo{author}{\bibfnamefont{D.}~\bibnamefont{K\"odderitzsch}},
  \bibnamefont{and} \bibinfo{author}{\bibfnamefont{P.~J.} \bibnamefont{Kelly}},
  \bibinfo{journal}{Phys. Rev. Lett.} \textbf{\bibinfo{volume}{107}},
  \bibinfo{pages}{066603} (\bibinfo{year}{2011}).

\bibitem[{\citenamefont{Sakuma}(2012)}]{Sak12}
\bibinfo{author}{\bibfnamefont{A.}~\bibnamefont{Sakuma}}, \bibinfo{journal}{J.
  Phys. Soc. Jpn.} \textbf{\bibinfo{volume}{81}}, \bibinfo{pages}{084701}
  (\bibinfo{year}{2012}).

\bibitem[{\citenamefont{Elze}(2012)}]{Elz12}
\bibinfo{author}{\bibfnamefont{H.-T.}~\bibnamefont{Elze}}, \bibinfo{journal}{Phys.
  Rev. A} \textbf{\bibinfo{volume}{85}}, \bibinfo{pages}{052109}
  (\bibinfo{year}{2012}).

\bibitem[{\citenamefont{Marx and Hutter}(2000)}]{MH00}
\bibinfo{author}{\bibfnamefont{D.}~\bibnamefont{Marx}} \bibnamefont{and}
  \bibinfo{author}{\bibfnamefont{J.}~\bibnamefont{Hutter}},
  \emph{\bibinfo{title}{Ab initio molecular dynamics: Theory and
  Implementation, {\rm In:} Modern Methods and Algorithms of Quantum
  Chemistry}}, NIC Series, Vol. 1, Ed. by J. Grotendorst, p. 301
  (\bibinfo{publisher}{John von Neumann Institute for Computing},
  \bibinfo{address}{J\"ulich}, \bibinfo{year}{2000}).

\bibitem{Daj14}
J. Dajka, Int.\ J.\ Theor.\ Phys.\ \textbf{53}, 870 (2014).

\bibitem{FLE14}
L. Fratino, A. Lampo, and H.-T. Elze, Phys. Scr. \textbf{T163}, 014005 (2014).

\bibitem[{\citenamefont{Mahani et~al.}(2014)\citenamefont{Mahani, Pertsova, and
  Canali}}]{MPC14}
\bibinfo{author}{\bibfnamefont{M.~R.} \bibnamefont{Mahani}},
  \bibinfo{author}{\bibfnamefont{A.}~\bibnamefont{Pertsova}}, \bibnamefont{and}
  \bibinfo{author}{\bibfnamefont{C.~M.} \bibnamefont{Canali}},
  \bibinfo{journal}{Phys. Rev. B} \textbf{\bibinfo{volume}{90}},
  \bibinfo{pages}{245406} (\bibinfo{year}{2014}).

\bibitem[{\citenamefont{Kondo}(1964)}]{Kon64}
\bibinfo{author}{\bibfnamefont{J.}~\bibnamefont{Kondo}},
  \bibinfo{journal}{Prog. Theor. Phys.} \textbf{\bibinfo{volume}{32}},
  \bibinfo{pages}{37} (\bibinfo{year}{1964}).

\bibitem[{\citenamefont{Hewson}(1993)}]{Hew93}
\bibinfo{author}{\bibfnamefont{A.~C.} \bibnamefont{Hewson}},
  \emph{\bibinfo{title}{The Kondo Problem to Heavy Fermions}}
  (\bibinfo{publisher}{Cambridge University Press},
  \bibinfo{address}{Cambridge}, \bibinfo{year}{1993}).

\bibitem{SGP12}
M.\ Sayad, D.\ G\"utersloh, and M.\ Potthoff,
Eur.\ Phys.\ J.\ B \textbf{85}, 125 (2012).

\bibitem{GL14}
J.\ P.\ Gauyacq and N.\ Lorente, 
Surf.\ Sci.\ \textbf{630}, 325 (2014).

\bibitem{DLZFR15}
F.\ Delgado, S.\ Loth, M.\ Zielinski, and J.\ Fern\'andez-Rossier, 
Europhys.\ Lett.\ \textbf{109}, 57001 (2015).

\bibitem{SRP15}
M.\ Sayad, R.\ Rausch, and M.\ Potthoff (to be published).

\bibitem[{\citenamefont{Wiesendanger}(2009)}]{Wie09}
\bibinfo{author}{\bibfnamefont{R.}~\bibnamefont{Wiesendanger}},
  \bibinfo{journal}{Rev. Mod. Phys.} \textbf{\bibinfo{volume}{81}},
  \bibinfo{pages}{1495} (\bibinfo{year}{2009}).

\bibitem[{\citenamefont{Nunes and Freeman}(1993)}]{NF93}
\bibinfo{author}{\bibfnamefont{G.}~\bibnamefont{Nunes}} \bibnamefont{and}
  \bibinfo{author}{\bibfnamefont{M.~R.} \bibnamefont{Freeman}},
  \bibinfo{journal}{Science} \textbf{\bibinfo{volume}{262}},
  \bibinfo{pages}{1029} (\bibinfo{year}{1993})..

\bibitem[{\citenamefont{Loth et~al.}(2010)\citenamefont{Loth, Etzkorn, Lutz,
  Eigler, and Heinrich}}]{LEL+10}
\bibinfo{author}{\bibfnamefont{S.}~\bibnamefont{Loth}},
  \bibinfo{author}{\bibfnamefont{M.}~\bibnamefont{Etzkorn}},
  \bibinfo{author}{\bibfnamefont{C.~P.} \bibnamefont{Lutz}},
  \bibinfo{author}{\bibfnamefont{D.~M.} \bibnamefont{Eigler}},
  \bibnamefont{and} \bibinfo{author}{\bibfnamefont{A.~J.}
  \bibnamefont{Heinrich}}, \bibinfo{journal}{Science}
  \textbf{\bibinfo{volume}{329}}, \bibinfo{pages}{1628} (\bibinfo{year}{2010}).

\bibitem[{\citenamefont{Morgenstern}(2010)}]{Mor10}
\bibinfo{author}{\bibfnamefont{M.}~\bibnamefont{Morgenstern}},
  \bibinfo{journal}{Science} \textbf{\bibinfo{volume}{329}},
  \bibinfo{pages}{1609} (\bibinfo{year}{2010}).

\bibitem[{\citenamefont{Wolf et~al.}(2001)\citenamefont{Wolf, Awschalom,
  Buhrman, Daughton, von Molnar, Roukes, Chtchelkanova, and Treger}}]{WAB01}
\bibinfo{author}{\bibfnamefont{S.~A.} \bibnamefont{Wolf}},
  \bibinfo{author}{\bibfnamefont{D.~D.} \bibnamefont{Awschalom}},
  \bibinfo{author}{\bibfnamefont{R.~A.} \bibnamefont{Buhrman}},
  \bibinfo{author}{\bibfnamefont{J.~M.} \bibnamefont{Daughton}},
  \bibinfo{author}{\bibfnamefont{S.}~\bibnamefont{von Molnar}},
  \bibinfo{author}{\bibfnamefont{M.~L.} \bibnamefont{Roukes}},
  \bibinfo{author}{\bibfnamefont{A.~Y.} \bibnamefont{Chtchelkanova}},
  \bibnamefont{and} \bibinfo{author}{\bibfnamefont{D.~M.}
  \bibnamefont{Treger}}, \bibinfo{journal}{Science}
  \textbf{\bibinfo{volume}{294}}, \bibinfo{pages}{1488} (\bibinfo{year}{2001}).

\bibitem[{\citenamefont{Khajetoorians et~al.}(2011)\citenamefont{Khajetoorians,
  Wiebe, Chilian, and Wiesendanger}}]{KWCW11}
\bibinfo{author}{\bibfnamefont{A.~A.} \bibnamefont{Khajetoorians}},
  \bibinfo{author}{\bibfnamefont{J.}~\bibnamefont{Wiebe}},
  \bibinfo{author}{\bibfnamefont{B.}~\bibnamefont{Chilian}}, \bibnamefont{and}
  \bibinfo{author}{\bibfnamefont{R.}~\bibnamefont{Wiesendanger}},
  \bibinfo{journal}{Science} \textbf{\bibinfo{volume}{332}},
  \bibinfo{pages}{1062} (\bibinfo{year}{2011}).

\bibitem[{\citenamefont{Scazza et~al.}(2014)\citenamefont{Scazza, Hofrichter,
  H\"ofer, Groot, Bloch, and Fölling}}]{SHH+14}
\bibinfo{author}{\bibfnamefont{F.}~\bibnamefont{Scazza}},
  \bibinfo{author}{\bibfnamefont{C.}~\bibnamefont{Hofrichter}},
  \bibinfo{author}{\bibfnamefont{M.}~\bibnamefont{H\"ofer}},
  \bibinfo{author}{\bibfnamefont{P.~C.~D.} \bibnamefont{Groot}},
  \bibinfo{author}{\bibfnamefont{I.}~\bibnamefont{Bloch}}, \bibnamefont{and}
  \bibinfo{author}{\bibfnamefont{S.}~\bibnamefont{Fölling}},
  \bibinfo{journal}{Nature Physics} \textbf{\bibinfo{volume}{10}},
  \bibinfo{pages}{779} (\bibinfo{year}{2014}).

\bibitem[{\citenamefont{Cappellini et~al.}(2014)\citenamefont{Cappellini,
  Mancini, Pagano, Lombardi, Livi, de~Cumis, Cancio, Pizzocaro, Calonico, Levi
  et~al.}}]{CMP+14}
\bibinfo{author}{\bibfnamefont{G.}~\bibnamefont{Cappellini}},
  \bibinfo{author}{\bibfnamefont{M.}~\bibnamefont{Mancini}},
  \bibinfo{author}{\bibfnamefont{G.}~\bibnamefont{Pagano}},
  \bibinfo{author}{\bibfnamefont{P.}~\bibnamefont{Lombardi}},
  \bibinfo{author}{\bibfnamefont{L.}~\bibnamefont{Livi}},
  \bibinfo{author}{\bibfnamefont{M.~S.} \bibnamefont{de~Cumis}},
  \bibinfo{author}{\bibfnamefont{P.}~\bibnamefont{Cancio}},
  \bibinfo{author}{\bibfnamefont{M.}~\bibnamefont{Pizzocaro}},
  \bibinfo{author}{\bibfnamefont{D.}~\bibnamefont{Calonico}},
  \bibinfo{author}{\bibfnamefont{F.}~\bibnamefont{Levi}}, \bibnamefont{et~al.},
  \bibinfo{journal}{Phys. Rev. Lett.} \textbf{\bibinfo{volume}{113}},
  \bibinfo{pages}{120402} (\bibinfo{year}{2014}).

\bibitem[{\citenamefont{Heslot}(1985)}]{Hes85}
\bibinfo{author}{\bibfnamefont{A.}~\bibnamefont{Heslot}},
  \bibinfo{journal}{Phys. Rev. D} \textbf{\bibinfo{volume}{31}},
  \bibinfo{pages}{1341} (\bibinfo{year}{1985}).

\bibitem[{\citenamefont{Hall}(2008)}]{Hal08}
\bibinfo{author}{\bibfnamefont{M.~J.~W.} \bibnamefont{Hall}},
  \bibinfo{journal}{Phys. Rev. A} \textbf{\bibinfo{volume}{78}},
  \bibinfo{pages}{042104} (\bibinfo{year}{2008}).

\bibitem[{\citenamefont{Yang and Hirschfelder}(1980)}]{YH80}
\bibinfo{author}{\bibfnamefont{K.-H.} \bibnamefont{Yang}} \bibnamefont{and}
  \bibinfo{author}{\bibfnamefont{J.~O.} \bibnamefont{Hirschfelder}},
  \bibinfo{journal}{Phys. Rev. A} \textbf{\bibinfo{volume}{22}},
  \bibinfo{pages}{1814} (\bibinfo{year}{1980}).

\bibitem[{\citenamefont{Lakshmanan and Daniel}(1983)}]{LD83}
\bibinfo{author}{\bibfnamefont{M.}~\bibnamefont{Lakshmanan}} \bibnamefont{and}
  \bibinfo{author}{\bibfnamefont{M.}~\bibnamefont{Daniel}},
  \bibinfo{journal}{J. Chem. Phys.} \textbf{\bibinfo{volume}{78}},
  \bibinfo{pages}{7505} (\bibinfo{year}{1983}).

\bibitem[{\citenamefont{Verner}(2010)}]{JHV2010}
\bibinfo{author}{\bibfnamefont{J.~H.} \bibnamefont{Verner}},
  \bibinfo{journal}{Numerical Algorithms} \textbf{\bibinfo{volume}{53}},
  \bibinfo{pages}{383} (\bibinfo{year}{2010}).

\bibitem[{\citenamefont{Lieb and Robinson}(1972)}]{LR72}
\bibinfo{author}{\bibfnamefont{E.~H.} \bibnamefont{Lieb}} \bibnamefont{and}
  \bibinfo{author}{\bibfnamefont{D.~W.} \bibnamefont{Robinson}},
  \bibinfo{journal}{Commun. Math. Phys.} \textbf{\bibinfo{volume}{28}},
  \bibinfo{pages}{251} (\bibinfo{year}{1972}).

\bibitem[{\citenamefont{Bravyi et~al.}(2006)\citenamefont{Bravyi, Hastings, and
  Verstraete}}]{BHV06}
\bibinfo{author}{\bibfnamefont{S.}~\bibnamefont{Bravyi}},
  \bibinfo{author}{\bibfnamefont{M.~B.} \bibnamefont{Hastings}},
  \bibnamefont{and}
  \bibinfo{author}{\bibfnamefont{F.}~\bibnamefont{Verstraete}},
  \bibinfo{journal}{Phys. Rev. Lett.} \textbf{\bibinfo{volume}{97}},
  \bibinfo{pages}{050401} (\bibinfo{year}{2006}).

\bibitem[{\citenamefont{Kikuchi}(1956)}]{Kik56}
\bibinfo{author}{\bibfnamefont{R.}~\bibnamefont{Kikuchi}}, \bibinfo{journal}{J.
  Appl. Phys.} \textbf{\bibinfo{volume}{27}}, \bibinfo{pages}{1352}
  (\bibinfo{year}{1956}).

\bibitem[{\citenamefont{Zhang and Li}(2004)}]{ZL04}
\bibinfo{author}{\bibfnamefont{S.}~\bibnamefont{Zhang}} \bibnamefont{and}
  \bibinfo{author}{\bibfnamefont{Z.}~\bibnamefont{Li}}, \bibinfo{journal}{Phys.
  Rev. Lett.} \textbf{\bibinfo{volume}{93}}, \bibinfo{pages}{127204}
  (\bibinfo{year}{2004}).

\bibitem[{\citenamefont{Fetter and Walecka}(1971)}]{FW71}
\bibinfo{author}{\bibfnamefont{A.~L.} \bibnamefont{Fetter}} \bibnamefont{and}
  \bibinfo{author}{\bibfnamefont{J.~D.} \bibnamefont{Walecka}},
  \emph{\bibinfo{title}{Quantum Theory of Many-Particle Systems}}
  (\bibinfo{publisher}{McGraw-Hill}, \bibinfo{address}{New York},
  \bibinfo{year}{1971}).

\bibitem[{\citenamefont{Balzer and Potthoff}(2011)}]{BP11}
\bibinfo{author}{\bibfnamefont{M.}~\bibnamefont{Balzer}} \bibnamefont{and}
  \bibinfo{author}{\bibfnamefont{M.}~\bibnamefont{Potthoff}},
  \bibinfo{journal}{Phys. Rev. B} \textbf{\bibinfo{volume}{83}},
  \bibinfo{pages}{195132} (\bibinfo{year}{2011}).

\bibitem[{\citenamefont{Press et~al.}(2007)\citenamefont{Press, Teukolsky,
  Vetterling, and Flannery}}]{PTVF07}
\bibinfo{author}{\bibfnamefont{W.}~\bibnamefont{Press}},
  \bibinfo{author}{\bibfnamefont{S.~A.} \bibnamefont{Teukolsky}},
  \bibinfo{author}{\bibfnamefont{W.~T.} \bibnamefont{Vetterling}},
  \bibnamefont{and} \bibinfo{author}{\bibfnamefont{B.}~\bibnamefont{Flannery}},
  \emph{\bibinfo{title}{Numerical Recipes}} (\bibinfo{publisher}{Cambridge},
  \bibinfo{address}{Cambridge}, \bibinfo{year}{2007}), \bibinfo{edition}{3rd}
  ed.

\bibitem[{\citenamefont{Fransson}(2008)}]{Fra08}
\bibinfo{author}{\bibfnamefont{J.}~\bibnamefont{Fransson}},
  \bibinfo{journal}{Nanotechnology} \textbf{\bibinfo{volume}{19}},
  \bibinfo{pages}{285714} (\bibinfo{year}{2008}).

\bibitem[{\citenamefont{Simanek and Heinrich}(2003)}]{SH03}
\bibinfo{author}{\bibfnamefont{E.}~\bibnamefont{Simanek}} \bibnamefont{and}
  \bibinfo{author}{\bibfnamefont{B.}~\bibnamefont{Heinrich}},
  \bibinfo{journal}{Phys. Rev. B} \textbf{\bibinfo{volume}{67}},
  \bibinfo{pages}{144418} (\bibinfo{year}{2003}).

\bibitem[{\citenamefont{Kambersk\'y}(2007)}]{Kam07}
\bibinfo{author}{\bibfnamefont{V.}~\bibnamefont{Kambersk\'y}},
  \bibinfo{journal}{Phys. Rev. B} \textbf{\bibinfo{volume}{76}},
  \bibinfo{pages}{134416} (\bibinfo{year}{2007}).

\bibitem[{\citenamefont{Brataas et~al.}(2008)\citenamefont{Brataas,
  Tserkovnyak, and Bauer}}]{BTB08}
\bibinfo{author}{\bibfnamefont{A.}~\bibnamefont{Brataas}},
  \bibinfo{author}{\bibfnamefont{Y.}~\bibnamefont{Tserkovnyak}},
  \bibnamefont{and} \bibinfo{author}{\bibfnamefont{G.~E.~W.}
  \bibnamefont{Bauer}}, \bibinfo{journal}{Phys. Rev. Lett.}
  \textbf{\bibinfo{volume}{101}}, \bibinfo{pages}{037207}
  (\bibinfo{year}{2008}).

\bibitem[{\citenamefont{Mankovsky et~al.}(2013)\citenamefont{Mankovsky,
  K\"odderitzsch, Woltersdorf, and Ebert}}]{MKWE13}
\bibinfo{author}{\bibfnamefont{S.}~\bibnamefont{Mankovsky}},
  \bibinfo{author}{\bibfnamefont{D.}~\bibnamefont{K\"odderitzsch}},
  \bibinfo{author}{\bibfnamefont{G.}~\bibnamefont{Woltersdorf}},
  \bibnamefont{and} \bibinfo{author}{\bibfnamefont{H.}~\bibnamefont{Ebert}},
  \bibinfo{journal}{Phys. Rev. B} \textbf{\bibinfo{volume}{87}},
  \bibinfo{pages}{014430} (\bibinfo{year}{2013}).

\bibitem[{\citenamefont{Thonig and Henk}(2014)}]{TH14}
\bibinfo{author}{\bibfnamefont{D.}~\bibnamefont{Thonig}} \bibnamefont{and}
  \bibinfo{author}{\bibfnamefont{J.}~\bibnamefont{Henk}}, \bibinfo{journal}{New
  J. Phys.} \textbf{\bibinfo{volume}{16}}, \bibinfo{pages}{013032}
  (\bibinfo{year}{2014}).

\bibitem[{\citenamefont{Ashcroft and Mermin}(1976)}]{AM76}
\bibinfo{author}{\bibfnamefont{N.~W.} \bibnamefont{Ashcroft}} \bibnamefont{and}
  \bibinfo{author}{\bibfnamefont{N.~D.} \bibnamefont{Mermin}},
  \emph{\bibinfo{title}{Solid State Physics}} (\bibinfo{publisher}{Holt,
  Rinehart and Winston}, \bibinfo{address}{New York}, \bibinfo{year}{1976}).

\end{thebibliography}
\end{document}